# The Global Minimum Tax, Investment Incentives and Asymmetric Tax Competition*

Xuyang Chen† Rui Sun‡

November 21, 2025

**Abstract**

This paper investigates the OECD's global minimum tax (GMT) in a formal model of tax competition between asymmetric countries. We consider both profit shifting and real responses of multinational enterprises, and highlight the role of the substance-based income exclusion (SBIE) in investment incentives and tax rate setting. The GMT reduces the true tax rate differential and benefits the large country, while the revenue effect is generally ambiguous for the small country. In the short run where tax rates are fixed, the GMT reduces the small country's revenue if profit shifting costs are low and increases it otherwise. In the long run where countries adjust tax rates, the GMT reshapes the tax game and the competition pattern. We reveal that the minimum rate binds the small country only if it is low. With the rise of the GMT rate, countries will set tax rates below the minimum to boost capital investments and collect top-up taxes. Simulations show that a moderate GMT rate can raise both countries' revenues and the large country's welfare in the long run. However, it may reduce the small country's welfare if the welfare weight of private income is high.

---

*We would like to thank Alan Auerbach, Chunyang Fu, Andreas Haufler, Jean Hindriks, Mathieu Parenti, Hylke Vandenbussche, Gonzague Vannoorenberghe, Wenjing Wang, Gabriel Zucman and seminar participants at UC Berkeley, Central University of Finance and Economics, and Zhongnan University of Economics and Law for helpful comments and suggestions on earlier drafts of this paper. Xuyang Chen gratefully acknowledges financial support from ARC (4110S000001) and FNRS (MIS 4129S000012).

†Zhongnan University of Economics and Law, UCLouvain & ULB. E-mail: xuyangchen@zuel.edu.cn.

‡Haas School of Business, University of California, Berkeley. E-mail: ruisun233@berkeley.edu.

# 1 Introduction

Multinational enterprises (MNEs) can exploit loopholes in tax rules to shift profits to low-tax countries to avoid paying taxes. International profit shifting has caused considerable losses of tax revenue for both OECD and developing countries.[1] To relieve the pressure on the outdated international corporate tax system, over 135 jurisdictions in 2021 agreed on a 15% global minimum tax (GMT), which is a key part of Pillar Two of the two-pillar solution proposed by the OECD/G20 Inclusive Framework on base erosion and profit shifting (BEPS). The GMT ensures that large MNEs with revenues above €750 million are subject to a 15% minimum tax rate in each jurisdiction where they operate.

In practice, the GMT under Pillar Two works in a roundabout way. The first step is to determine the jurisdictional effective tax rate (ETR) of the MNE. It is computed by dividing the taxes (called covered taxes) of the affiliate in that jurisdiction by the income (called GloBE income) it has. If the jurisdictional ETR is below 15%, the MNE will be subject to a top-up tax. The top-up tax rate is the difference between the 15% minimum rate and the ETR. The top-up tax base (called excess profit) is calculated as the GloBE income in excess of the Substance-Based Income Exclusion (SBIE). The SBIE allows MNEs to tax-deduct a percentage of the carrying value of tangible assets and payroll expenses from the GloBE income of the low-tax affiliate. The current carve-out rate is 8% on tangible assets and 10% on payroll costs, while it will reduce to 5% on both tangibles and payroll over a transition period of ten years (see OECD, 2021; European Commission, 2021; Devereux et al., 2021).[2] Hence, the SBIE benefits the low-tax affiliates that have real economic activity by reducing their top-up tax liability. A key question is which countries receive the additional tax revenue. In principle, the top-up tax can be collected either by the country where the headquarters of the MNE resides or by the host country where the MNE's affiliate records the profit. The first scenario corresponds to the income inclusion rule (IIR), while the latter corresponds to the qualified domestic minimum top-up tax (QDMTT). The Model Rules published by the OECD/G20 Inclusive Framework in December 2021 introduced the QDMTT, which gives the host country the priority to collect top-up taxes over the headquarters country.[3]

[1] For instance, Crivelli et al. (2016) argue that developing countries suffer from MNEs' profit shifting to a greater extent than OECD countries. In their estimates, the revenue loss is 1.3% of GDP for developing countries, compared to 1% for advanced economies. Bilicka (2019) shows that foreign multinational subsidiaries in the UK underreport taxable profits by about 50 percent relative to domestic companies. The potential tax revenue gains from closing the gap would be £3 billion in 2000 and £25 billion in 2014. Davies et al. (2018) show that transfer pricing by manufacturing firms causes France the tax revenue loss of €333 million in 1999, amounting to 1% of its total corporate tax revenue that year. See also Wier and Zucman (2022) and Tørsløv et al. (2023) for comprehensive analyses of profit shifting and the associated revenue loss worldwide.

[2] Specifically, during the transition period, the carve-out rate will decline annually by 0.2 percentage points in the first five years, and by 0.4 percentage points for tangible assets and by 0.8 percentage points for payroll in the last five years.

In this paper, we investigate the OECD's GMT in a formal model of international tax competition with two *asymmetric* countries that differ in market size. First, two countries choose corporate tax rates noncooperatively to maximize their revenues. Then the MNE chooses capital investment in each country and profit shifting level to maximize total after-tax profits, taking as given the tax environments. Following the previous literature on minimum taxation (e.g., Kanbur and Keen, 1993; Keen and Konrad, 2013), we focus on the situation where the minimum rate lies between two countries' initial tax rates without the GMT. Our model captures the key features of Pillar Two (i.e., the SBIE and the QDMTT), and takes into account both profit shifting and real responses of the MNE. We analyze how the GMT affects the MNE's behavior and countries' taxes from a short-run perspective where corporate tax rates are fixed, and from a long-run perspective where governments engage in tax competition. This approach disentangles the tax incentive effect of the GMT due to the tax deduction of the SBIE from the behavioral adjustments by the governments.

We start from the short-run analysis, in which only the MNE can adjust its behavior. Introducing the GMT does not change the investment level in the large country. With the top-up tax paid by the low-tax affiliate, the GMT reduces the difference between tax rates on two affiliates' GloBE incomes (referred to as the *true* tax rate differential). Consequently, profit shifting is reduced, and the large country's tax revenue definitely increases. However, the revenue effect of the GMT is generally ambiguous for the small country. Starting from the equilibrium without the GMT, a marginal increase in the minimum rate has two opposite effects. Firstly, due to the SBIE, a higher GMT rate exerts investment incentives for the low-tax affiliate and thus raises the small country's revenue. Secondly, a higher GMT rate increases the top-up tax rate and incurs a larger revenue loss from the deduction of SBIE. Which effect can dominate depends on the initial tax rate of the small country (or equivalently, on the profit shifting cost of the MNE).

[3]In addition to the QDMTT and IIR, Pillar Two introduces the undertaxed profits rule (UTPR) as a backstop. In practice, the GMT under Pillar Two uses a hierarchical rule order, which specifies that the QDMTT is adopted first, followed by the IIR and the UTPR. In case that neither the QDMTT nor the IIR is applied, the UTPR allows jurisdictions to impose top-up taxes on local affiliates of the MNE to make up for any remaining low-taxed profits in the parent jurisdiction or any other third jurisdiction (see OECD, 2021; Devereux et al., 2022).

Then we investigate the long-run situation where both the MNE and the governments react to the tax reform. Countries compete in corporate tax rates to maximize their revenues while taking as given the international tax architecture.[4] At the equilibrium of the tax game, the GMT does not necessarily bind the small country. It is binding only if the minimum tax is low. Otherwise, the small country will set its tax rate below the minimum, even at zero (if the carve-out rate is relatively small), and collect the top-up tax.[5] The key insight is that due to tax deduction of the SBIE, lowering one country's tax rate below the GMT rate can incentivize real investment but also incurs a larger revenue loss since the top-up tax rate increases. When the GMT rate exceeds a certain threshold, the former effect can dominate. In particular, when the GMT rate is sufficiently high, the large country will also undercut the minimum so that profit shifting ends.[6] In this case, countries aim at attracting capital investments instead of competing for paper profits.

After characterizing the Nash equilibrium of the tax competition game, we examine the long-run revenue effects of the GMT. As in the short run, the GMT reduces the true tax rate differential between two countries and always benefits the large country. In contrast, the revenue effect for the small country is more subtle. We show that even a marginal GMT reform (with the minimum rate marginally above the small country's initial tax rate) may harm it. This happens when the asymmetry between two countries is small and the profit shifting cost is intermediate. The key insight is that in this case the marginal reform induces both countries to undercut the minimum so that the profit shifted to the small country jumps *discontinuously* to zero. Since the small country attracts considerable paper profits absent the GMT, the revenue loss from eliminating profit shifting can outweigh the increased taxation on true profits generated by real economic activities. Furthermore, we provide sufficient conditions for the Pareto-improving non-marginal tax reform. When the carve-out rate is not very small, the GMT rate is not very high and profit shifting is not sensitive to the minimum, the GMT can benefit the small country by increasing the taxation of both true profit and shifted profit.

[4]The GMT rate is currently set at 15%, but this decision has provoked substantial debates. Many observers argue that a minimum rate of 15% is too low to effectively curb MNEs' profit shifting and advocate for a much higher minimum rate. For instance, the EU Tax Observatory and Oxfam have called for a tax floor of 21% or even 25% to boost corporate tax revenues for developed and low-income countries (see https://fortune.com/2021/06/07/global-minimum-corporate-tax-rate-g7-15-percent-pact/). Notably, the Biden administration initially proposed a 21% minimum tax rate as part of its $2 trillion infrastructure spending plan (see https://bidenwhitehouse.archives.gov/briefing-room/statements-releases/2021/03/31/fact-sheet-the-american-jobs-plan/), while the rate was lowered to 15% at the OECD meetings in order to gain broad support from low-tax countries. In light of this, our analysis is not restricted to the 15% minimum rate. Instead, we investigate the GMT in a more general setting and show the impact of different minimum rates on countries' equilibrium taxes.

[5]Remarkably, some countries (e.g., Bulgaria, Hungary, Ireland and the United Arab Emirates) maintain their corporate tax rates below 15% and adopt the QDMTT to top up the rate to 15% for local affiliates of large MNEs (see PwC's survey on Pillar Two implementation, available at https://taxsummaries.pwc.com/). This provides preliminary evidence that introducing the GMT may not raise all countries' equilibrium tax rates above the minimum rate.

[6]In our paper, "undercutting the minimum" means that "setting the tax rate below the minimum rate and collecting top-up taxes". We use the two expressions interchangeably.

We then extend our analysis to the case where countries maximize their welfare — which is a combination of the tax revenue and private income (i.e., the MNE's after-tax profit). Numerical simulations show that: the small country increases its tax to the minimum rate if the welfare weight of private income is low, and undercuts the minimum otherwise. Intuitively, lowering one country's tax rate below the minimum can increase the MNE's after-tax profit through the deduction of the SBIE. When the welfare weight of private income rises, in addition to the trade-off between incentivizing capital investment and incurring tax revenue loss, the effect of the tax rate reduction on private income is stronger, which induces the small country to undercut the minimum rate. Our results indicate that equilibrium tax revenues of both countries increase under a moderate GMT rate of 15%. Besides, we establish that introducing the GMT improves the large country's welfare if the sum of the cross-border tax base elasticity and the own tax base elasticity (in absolute value) is larger than unity. On the other hand, the GMT may reduce the small country's welfare when the welfare weight of private income is high. The reason is that in this case, the loss of private income would outweigh the tax revenue gain for the small country. Based on our results, we discuss the policy implications relating to current status of Pillar Two implementation and previous anti-avoidance measures (e.g., the OECD/G20 BEPS Project).

Our paper contributes to the emerging theoretical works on the GMT. Firstly, we take into account the real investment responses of the MNE. In most theoretical GMT literature, firms' profits are assumed to be fixed and independent of tax rates so that only profit shifting behavior is considered. This assumption implies the adoption of a pure profit tax with full tax deductibility of costs, which is at odds with most countries' corporate tax systems. In contrast, our model allows for the partial deductibility of capital costs such that countries' tax policies affect the MNE's decision on both profit shifting and real investment.[7] As is discussed below, partial deductibility creates scope for countries to gain tax revenue by incentivizing affiliates to increase investments. Secondly, we capture the SBIE and QDMTT in the tax competition game. The SBIE is a key feature of Pillar Two but is largely ignored in the existing theoretical literature. Only a few papers theoretically investigate the effects of the SBIE (see Devereux et al., 2021, 2022; Schjelderup and Stähler, 2024), whereas the strategic interactions between countries' tax rates are overlooked in these works. When capital costs are not fully deductible, due to tax deduction of the SBIE, either increasing the minimum rate in the short run or setting tax rate below the minimum in the long run can exert investment incentives, while doing so also increases the top-up tax rate and incurs a larger revenue loss (see Sections 4 and 5 for details). This is the key trade-off in our analysis, which determines the short-run revenue effect of the GMT and reshapes the tax competition game in the long run. Thirdly, our results are comparable with traditional minimum taxation literature. When all costs are tax deductible (which is equivalent to the assumption of fixed profits), investments in each country are undistorted at the initial equilibrium. The aforementioned tax policies fail to raise the affiliate's GloBE income through investment incentives, but only causes a larger revenue loss. Consequently, the GMT always harms the small country in the short run. In the long run, undercutting the minimum is a strictly dominated strategy for each country. Therefore, the GMT under Pillar Two works in the same way as "a constraint that no tax rate may be set below the minimum level". In this sense, the traditional minimum taxation model can be regarded as a special case in our paper.

The rest of the paper proceeds as follows. Section 2 briefly discusses the related literature. In Section 3, we present the model and analyze the equilibrium before the GMT is introduced. Section 4 analyzes the short-run effects of the GMT. In Section 5, we characterize the Nash equilibrium of the tax competition game and investigate the long-run effects of the GMT. In Section 6, we consider an extension in which countries maximize their welfare.

[7] In this regard, our paper is related to Chen and Hindriks (2023), who analyze the effects of tax deductibility in a model of tax competition with two countries and a tax haven. They derive conditions under which pure profit tax is superior (inferior) to turnover tax and investigate the optimal deductibility rate in the profit shifting context. In our paper, the deductibility rate is fixed and our focus is on the effects of the GMT rate and carve-out rate.

Section 7 concludes.

## 2 Related literature

A number of recent studies empirically analyze the tax revenue (or welfare) consequences of the GMT in the context of fixed corporate tax rates. UNCTAD (2022) shows that the GMT (with the SBIE) could lead to a growth in global tax revenues generated by FDI income between 15% and 20%, if all host countries apply the QDMTT. Baraké et al. (2022) document that G7 countries could collect around €90 billion in the headquarters scenario, while it would fall to €17 billion under host country collection. By contrast, developing countries would favor the QDMTT over the IIR. Ferrari et al. (2023) model and quantify the effects of the GMT, showing that it can improve welfare in most countries by inducing higher tax revenues. Hugger et al. (2024) find that the GMT would narrow the tax differential and raise global tax revenues by between \$155 and \$192 billion on average per year. Very few empirical works take into account the tax rate adjustments by countries in response to the minimum tax. IMF (2023) estimate that the average tax rate would rise from 22.2% to 24.3% in response to the GMT, which in turn could increase global tax revenues by 8.1%. Moreover, a growing empirical literature investigates various minimum tax reforms implemented at the national level (e.g., Buettner and Poehnlein, 2024; Lobel et al., 2024; Bukovina et al., 2025). These minimum taxes are unilateral measures mainly targeting domestic firms in a single-country context.[8] In contrast, the OECD's GMT applies to large MNEs and represents a form of tax coordination in international settings.

Our paper is related to the traditional minimum taxation literature, which treats the minimum tax as a lower bound imposed on countries' tax rates. Kanbur and Keen (1993) initiatively explore the minimum tax in a commodity tax competition model with cross-border shopping. They show that both large and small countries can benefit from the minimum tax. More recently, Hebous and Keen (2023) extend this framework to study international taxation of MNEs and derive the levels of maximal Pareto dominant minimum tax rate and Pareto-efficient minimum rate. In the two papers, the minimum tax binds the small country and induces the large country to set tax rate along the unconstrained best response curve. Wang (1999) extends Kanbur and Keen's (1993) model to the Stackelberg tax-setting, and presents that the minimum tax not only binds the follower (i.e., the small country) but also may bind the leader (i.e., the large country). Besides, imposing a minimum tax constraint definitely benefits the leader and harms the follower. In our paper, we restrict attention to simultaneous move of countries, as is widely accepted in the profit shifting literature. The binding minimum rate is also an implicit assumption in Janeba and Schjelderup (2023), who study the GMT in a model where two identical non-haven countries compete for firms via tax rates or subsidies, with profits shifted to the tax haven. They assume that the haven's tax rate adjusts to the minimum rate once the GMT is introduced, and show that the revenue effects of the GMT are ambiguous, depending on the fiscal instrument governments use. However, specifying the minimum tax as a constraint that no tax rate may be set below a minimum level somewhat deviates from the design of the GMT based on the OECD's Model Rules. The GMT under Pillar Two allows countries to collect additional revenue via top-up taxes when an affiliate's ETR falls below the minimum rate. Indeed, some countries keep their tax rates below the minimum and impose the QDMTT on affiliates operating in their territories (see footnote 5). In particular, the Irish Department of Finance states that "the positive effects will be greater than the challenges, as the agreement (...) provide(ing) a sound and stable basis for inward investment into Ireland in the long-term".[9] This is consistent with the argument in our paper that countries may undercut the GMT rate to attract inward investments.

[8] Specifically, Lobel et al. (2024) study a minimum tax introduced in Honduras between 2014 and 2017 – which required firms with gross revenue above a threshold to pay the maximum between their liability under profit tax and 1.5 percent of gross revenue. Bukovina et al. (2025) evaluate the 2014 minimum tax reform in Slovakia, which prescribed fixed minimum tax payments while allowing carryforwards of minimum tax credits. Buettner and Poehnlein (2024) analyze a minimum tax introduced in Germany in 2004 that forced German municipalities to set tax rates of at least 9.1%.

A few recent theoretical papers on the GMT go beyond "the minimum tax constraint assumption" and consider the case where countries' tax rates are below the GMT rate. Johannesen (2022) sets up a model of tax competition for paper profits among tax havens and non-haven countries, and assumes that the top-up taxes are collected by home countries (under the IIR). The GMT causes a loss of private consumption for the owners of the multinationals in non-haven countries, but also curbs profit shifting and boosts tax revenue. The net welfare effect is generally ambiguous for non-haven countries. Haufler and Kato

[9]Information is available at: https://www.gov.ie/en/department-of-finance/press-releases/minister-mcgrath-notes-irelands-application-of-effective-15-corporation-tax-rate-for-in-scope-businesses/.

(2024) develop a tax competition model, where a non-haven and a haven country are bound by the GMT rate for large MNEs, but can choose tax rates freely for small MNEs. They show that introducing a moderate minimum tax can raise tax revenues for both countries. As the GMT rate increases, each country has incentives to split the tax rate and set tax rate below the minimum for small MNEs. However, the SBIE is omitted in the two papers. Schjelderup and Stähler (2024) consider the SBIE in a standard MNE model where the host country's tax rate is below the GMT rate. They show that the SBIE works like a wage subsidy and investment subsidy for the low-tax affiliate, since it allows the firm to deduct payroll costs and user costs of tangible assets twice from the overall tax base. However, in their model countries' tax rates are assumed to be exogenously given, and the analysis of the revenue effect of the GMT is absent.

Finally, distinguishing between the short-run and long-run perspectives relates our paper to the literature comparing two alternative tax principles – separate accounting (SA) and formula apportionment (FA) – in the taxation of MNEs (e.g., Riedel and Runkel, 2007; Mardan and Stimmelmayr, 2018).[10] Using a similar approach, these works evaluate the effects of switching from a system of SA to FA. In our paper, we restrict attention to SA, since it is the predominant tax principle at the international level.

## 3 The model

Consider two asymmetric countries, labelled by 1 and 2, that form a small part of the world. Each country hosts an affiliate of a representative multinational enterprise (MNE). Each affiliate produces a homogenous good according to a decreasing returns to scale technology $f_i(k_i)$, where $k_i$ is the capital employed by affiliate $i$. Decreasing returns to scale in production imply the existence of a fixed factor (e.g., entrepreneurial services) that generates economic rents. In Appendix E, we briefly present the extended model that includes labor. Output is sold at the world market at a price normalized to unity. The two affiliates pay the (exogenously given) world interest rate $r$ for per unit of capital use. For most corporate tax systems, the capital cost may not be fully tax deductible, since countries only allow the MNE to deduct the cost financed by debt but not by equity. Denote by $\mu \in [0,1)$ the fraction of capital cost that can be deducted from the corporate tax base. Although we mainly focus on the case of partial deductibility, our results also hold true for $\mu = 1$. As we will present in Sections 4 and 5, a pure profit tax with full deductibility – which is the underlying assumption in the existing minimum tax literature – can be regarded as a special case of our model. Since the main purpose of the paper is to investigate the effects of GMT, we impose a few structure on the production technology by specifying a quadratic production function: $f_i(k_i) := \alpha_i k_i - \frac{k_i^2}{2}$ with $\alpha_1 > \alpha_2 > r$.[11] With this set-up, country 1 (country 2) is the large country (small country) in the sense that country 1 has a lager market size and a less elastic tax base.

[10] Different from SA under which profit is taxed in the country where the MNE declares it, under FA the MNE's taxable incomes are consolidated first and then assigned to each country based on a formula. Some countries (e.g., the US, Canada and Germany) are applying the FA principle at the national level.

The MNE can shift profits between two affiliates in order to minimize its tax liability. We abstract from the specific channels through which the MNE reallocates profits and denote the profit shifting level by $g$. If $g > 0$ ($g < 0$), then the MNE shifts profit to (from) country 2 and so the tax base in country 2 goes up (down). Profit shifting is costly for the MNE and involves a non-deductible concealment cost with the form $h(g) := \frac{\delta}{2} g^2$, $\delta > 0$ . The concealment cost reflects that the MNC has to exert efforts – such as hiring lawyers and accountants – to defend its tax avoidance activities during audits by tax authorities (see Devereux et al., 2008; Mardan and Stimmelmayr, 2018, 2020; Koethenbuerger et al., 2019; Janeba and Schjelderup, 2023).[12] The tax rate of country $i$ ($i = 1, 2$) is denoted by $t_i$. The GloBE income (i.e., taxable profit) of affiliate $i$ is $\pi_i = f_i(k_i) - \mu r k_i + (-1)^i g$.

We consider a two-stage tax competition game. In the first stage, two countries choose tax rates simultaneously and non-cooperatively to maximize their own tax revenues. In the second stage, the MNE chooses real investment and profit shifting level to maximize total after-tax profits. The assumption of revenue-maximizing governments is frequently used in corporate tax competition models (e.g., Johannesen, 2010; Mardan and Stimmelmayr, 2018; Koethenbuerger et al., 2019; Janeba and Schjelderup, 2023; Haufler and Kato, 2024). It reflects the policy focus on increasing tax payments by MNEs and is line with the objective of the OECD/G20 Inclusive Framework on BEPS. In reality, profit shifting causes severe revenue shortfalls and raises equality-of-treatment concerns in many countries, which motivates politicians to stabilize the revenue from corporate taxes. In Section 6, we relax the assumption of revenue maximization and analyze the case of welfare-maximizing governments.

[11] The quadratic specification of a production function is usual in the profit shifting literature (e.g., Devereux et al., 2008; Johannesen, 2010; Haufler et al., 2018; Mardan and Stimmelmayr, 2020).

[12] We use the terms "concealment cost" and "profit shifting cost" interchangeably throughout the paper.

Without the GMT, the MNE's after-tax profit is:

$$\Pi = \sum_{i=1}^{2} \Big[ (1-t_i) \Big( f_i(k_i) - \mu r k_i + (-1)^i g \Big) - (1-\mu)\, r k_i \Big] - \frac{\delta}{2} g^2. \tag{1}$$

Solving the MNE's profit maximization leads to:

$$k_i = \max \left\{ \frac{\alpha_i (1-t_i) - (1-\mu t_i) r}{1-t_i}, 0 \right\}, \quad i = 1, 2, \tag{2}$$

$$g = (-1)^i \min \left\{ \frac{t_j - t_i}{\delta}, f_j(k_j) - \mu r k_j \right\} \text{ if } t_j \geqslant t_i. \tag{3}$$

From (2), for interior solution with $k_i > 0$, $\frac{\partial k_i}{\partial t_i} < 0$. Intuitively, a higher tax rate increases the tax burden of the affiliate and so discourages real investment. From (3), the MNE shifts profit from high-tax country to low-tax country. In particular, the high-tax affiliate reports zero taxable profit if the tax differential is sufficiently large. (2) and (3) replicate the standard results in the profit shifting literature.

The tax revenue function of country $i$ reads:

$$R_i(t_i, t_j) = t_i \pi_i = t_i \left( f_i(k_i) - \mu r k_i + (-1)^i g \right). \tag{4}$$

Due to the complexity inherent in partial tax deductibility and asymmetry settings, the model has no closed-form solution for all $\mu \in [0, 1)$. In what follows, we assume that $\frac{\alpha_1 - r}{\alpha_1 - \mu r} \leqslant 2 \frac{\alpha_2 - r}{\alpha_2 - \mu r}$ (or equivalently, $\underline{\alpha}_2 := \frac{r[\alpha_1(2-\mu) - \mu r]}{\alpha_1 + r - 2\mu r} \leqslant \alpha_2 < \alpha_1$). It means that country 2 is not "too small" relative to country 1. As shown in Appendix A.1, under this assumption country 2 has no incentive to become a tax haven with shifted profits being its only tax base.[13] Denote by $t_i^N$ country $i$'s equilibrium tax rate absent the GMT. Then we can present

[13] We do not consider tax havens with very small market-size and few genuine economic activities. Since the SBIE is not applicable to shell companies, the GMT will increase the tax rate on taxable profits of tax haven affiliates to 15% through levying a top-up tax. Introducing the GMT narrows the tax differential between non-haven countries and tax havens, thereby reducing profits shifted to the latter. This will hurt tax havens, since their success fundamentally relies on attracting substantial amounts of paper profits. Instead, our focus is on low-tax small countries with real economic activities (e.g., Ireland, Hungary and the UAE). As we will show, the effects of the GMT are non-trivial for these countries.

the following:

**Lemma 1.** *Before the introduction of the GMT,*

*(i) there exists a unique Nash equilibrium with* $t_i^N \in \left(0, \frac{\alpha_i - r}{\alpha_i - \mu r}\right)$, $i = 1, 2$;

*(ii) at equilibrium, the small country always undercuts the large country, i.e.,* $t_1^N > t_2^N$.

*Proof.* See Appendices A.1 and A.3. □

The lemma establishes the existence and uniqueness of the Nash equilibrium absent the GMT. There are capital investments in both countries at equilibrium, while the small country sets a lower tax rate to attract profits from the large country. Intuitively, under equal tax rates $t_1 = t_2$ the small country has a higher tax base elasticity (in absolute value) than the large country (i.e., $-\frac{\partial \pi_2}{\partial t_2}\frac{t_2}{\pi_2} > -\frac{\partial \pi_1}{\partial t_1}\frac{t_1}{\pi_1}$). Hence, the small country's tax base is more sensitive to tax rate changes, which forces it to tax less.

**Remark 1.** Using (A.4) and Lemma 1(ii) leads to $\frac{\partial t_1^N}{\partial \delta} > 0$, i.e., the equilibrium tax rate of country 1 strictly increases with the profit shifting cost. Denote by $\bar{t}_1 := \lim\limits_{\delta \to +\infty} t_1^N(\delta)$ the (least) upper bound for $t_1^N$. In contrast, the effect of profit shifting cost on country 2's equilibrium tax rate is ambiguous. Numerical simulations indicate that $t_2^N$ may *decrease* with $\delta$ when $\delta$ is large.[14]

**Remark 2.** When $\delta \to +\infty$, the profit shifting is eliminated so that the positive tax externality vanishes. In this limiting case, by setting the efficient tax rate $\bar{t}_1$, country 1 can achieve the first-best solution $\bar{R}_1 := \lim\limits_{\delta \to +\infty} R_1(t_1^N(\delta), t_2^N(\delta))$ absent the GMT. Formally, we have the following:

$$R_1(t_1^N, t_2^N) = \varphi_1(t_1^N) - \frac{t_1^N\left(t_1^N - t_2^N\right)}{\delta} < \varphi_1(t_1^N) < \varphi_1(\bar{t}_1) = \bar{R}_1, \tag{5}$$

where $\varphi_1(t_1^N) := t_1^N\left(f_1(k_1^N) - \mu r k_1^N\right)$ denotes the revenue from taxing affiliate 1's true profit generated by substantive activities.

[14] For example, consider the parameter specification: $\alpha_1 = 1.88$, $\alpha_2 = 1.3$, $r = 1$, $\mu = 0.1$. Then we have: $t_2^N = 0.1392$ when $\delta = 4$, and $t_2^N = 0.1390$ when $\delta = 7$.

(5) indicates that $\bar{R}_1$ is the upper bound for country 1's equilibrium revenue without the GMT. Nevertheless, after introducing the GMT, country 1 can attain an equilibrium revenue larger than $\bar{R}_1$ when the minimum is sufficiently high and the carve-out is not very small (see Section 5 and Appendix A.7.4).

# 4 Short-run analysis of the GMT: fixed tax rates

After introducing the GMT, affiliate $i$'s GloBE income $\pi_i$ is targeted for additional taxation (i.e., top-up tax) when country $i$'s tax rate falls below the GMT rate $t_m$. As in the previous minimum taxation literature (see Kanbur and Keen, 1993; Wang, 1999; Keen and Konrad, 2013), we assume throughout the paper that the GMT rate lies between the initial tax rates of two countries, i.e., $t_m \in (t_2^N, t_1^N)$.[15] The top-up tax rate for affiliate $i$ is $\max\{t_m - t_i, 0\}$, which is the difference between the minimum rate and the host country's tax rate. The SBIE allows the MNE to tax-deduct a fraction of the capital stock from the GloBE income, which reduces the top-up tax liability of the affiliate with substantive activity. The GloBE income after the deduction of the SBIE is the excess profit $E_i = \pi_i - \sigma k_i$, where $\sigma$ denotes the carve-out rate and $\sigma k_i$ is the SBIE. The excess profit constitutes the tax base for the top-up tax. So the top-up tax owed by affiliate $i$ is $\max\{t_m - t_i, 0\} \cdot (\pi_i - \sigma k_i)$. Notice that an MNE's tax liability is the same, no matter which country collects the top-up tax. The low-tax country has very strong incentives to implement the QDMTT, because otherwise it would leave "money on the table" for other countries (according to the IIR or the UTPR), without changing the MNE's tax burden.[16] In light of this, we assume throughout the paper that each country imposes the QDMTT on the affiliate that is recording undertaxed profits in its territory.

In the short run, two countries' tax rates remain unchanged, while the MNE is able to adjust the decisions on investment and profit shifting. This reflects the fact that governments usually need some time to adjust tax rates in response to the tax reform.

[15] Recall that the aim of Pillar Two is to increase tax payments by low-tax affiliates of large MNEs and deter aggressive tax avoidance. In our model, country 1 represents the high-tax large countries around the world. Introducing a GMT rate that is above the high-tax country's initial rate is unappealing to policymakers and is politically infeasible.

[16] See IMF (2023, pp. 24-25) and Perry (2023) for more explanations of why countries should adopt the QDMTT.

The after-tax profit of the MNE reads:

$$
\begin{aligned}
\Pi^m &= \sum_{i=1}^{2}\left[(1-t_i^N)\underbrace{\left(f_i-\mu r k_i+(-1)^i g\right)}_{\text{GloBE income}}-(1-\mu)rk_i\right]-\underbrace{(t_m-t_2^N)(f_2-\mu r k_2+g-\sigma k_2)}_{\text{top-up tax}}-\frac{\delta}{2}g^2\\
&=(1-t_1^N)(f_1-\mu r k_1-g)-(1-\mu)rk_1+(1-t_m)(f_2-\mu r k_2+g)-(1-\mu)rk_2\\
&\quad+\underbrace{(t_m-t_2^N)\sigma k_2}_{\text{tax saved from the SBIE}}-\frac{\delta}{2}g^2,
\end{aligned}
\tag{6}
$$

where superscript $m$ represents the introduction of the GMT.

As shown by (6), with the additional GloBE top-up tax, affiliate 2's GloBE income is now taxed at the minimum rate $t_m$. On the other hand, the SBIE reduces the exposure to the minimum tax so that the tax amount of $(t_m-t_2^N)\sigma k_2$ can be saved by affiliate 2.

The first-order conditions of the profit maximization are:

$$
\frac{\partial \Pi^m}{\partial k_1}=(1-t_1^N)(f_1'-\mu r)-(1-\mu)r=0 \implies k_1^m=\frac{\alpha_1\left(1-t_1^N\right)-(1-\mu t_1^N)r}{1-t_1^N}=k_1^N, \tag{7}
$$

$$
\begin{aligned}
\frac{\partial \Pi^m}{\partial k_2}&=(1-t_m)(f_2'-\mu r)-(1-\mu)r+\sigma(t_m-t_2^N)\leqslant 0 \implies\\
k_2^m&=\max\left\{\frac{\alpha_2(1-t_m)-(1-\mu t_m)r+(t_m-t_2^N)\sigma}{1-t_m},0\right\},
\end{aligned}
\tag{8}
$$

$$
\frac{\partial \Pi^m}{\partial g}=t_1^N-t_m-\delta g=0 \implies g^m=\frac{t_1^N-t_m}{\delta}. \tag{9}
$$

Notably, for a high carve-out rate, affiliate 2's excess profit $E_2$ will be negative so that the GMT is immaterial. Since the main focus of this paper is the effects of the GMT, we restrict attention to positive excess profit and assume that the short-run carve-out rate satisfies $\sigma \leqslant \frac{\alpha_2(1-t_m)+r(1-(2-t_m)\mu)}{2-t_2^N-t_m} =: \bar{\sigma}^S(t_m)$, where superscript $S$ represents the short-run analysis. As shown in Appendix B.1, this assumption ensures that $E_2>0$.

The GMT is inactive for affiliate 1, since country 1's tax rate is above the minimum. As shown by (7), the capital investment in country 1 is unaffected. In contrast, increasing the GMT rate has two opposite effects on the capital investment of affiliate 2. Specifically, for interior solution with $k_2^m > 0$, differentiating (8) with respect to $t_m$ yields:

$$\frac{\partial k_2^m}{\partial t_m} = \underbrace{\frac{\partial}{\partial t_m}\left[\frac{\alpha_2(1-t_m)-(1-\mu t_m)r}{1-t_m}\right]}_{\text{tax burden effect}} + \underbrace{\frac{\partial}{\partial t_m}\left[\frac{(t_m-t_2^N)\sigma}{1-t_m}\right]}_{\text{tax incentive effect}} = \frac{\sigma\left(1-t_2^N\right)-r\left(1-\mu\right)}{\left(1-t_m\right)^2}.$$

The first team on the right-hand side is negative. A higher GMT rate increases the taxation on affiliate 2's GloBE income and tends to reduces the real investment in country 2. We refer to it as the "tax burden effect". The second term is positive and captures the "tax incentive effect". A higher GMT rate means that more taxes can be saved due to the deduction of the SBIE (cf. (6)), which incentivizes affiliate 2 to employ more capital. Moreover, a higher carve-out rate makes the tax incentive effect stronger, since it leads to a larger SBIE. When $\sigma > \frac{r(1-\mu)}{1-t_2^N}$ ($\sigma < \frac{r(1-\mu)}{1-t_2^N}$), the second (first) effect dominates such that country 2's investment level increases (decreases) with the minimum rate.[17]

(9) indicates that the GMT reduces the profit shifting of the MNE. From (6), the MNE's profit shifting decision depends on the difference between the tax rates on two affiliates' GloBE incomes (taking into account the top-up tax paid by the low-tax affiliate). Henceforth, we refer to it as the *true* tax rate differential between two countries. After introducing the GMT, the tax rate on affiliate 2's GloBE income increases from $t_2^N$ to $t_m$. So the true tax differential narrows down and profit shifting decreases.

The short-run tax revenues of two countries are:

$$R_1^m = t_1^N(f_1 - \mu r k_1^m - g^m), \tag{10}$$

[17]Based on our results, we may examine the investment effects in real-world tax systems. We first estimate $\mu$ and $r$ à la Riedel and Runkel (2007). Assume that the interest rate is $\rho$ and that a share $\theta_\rho$ of the capital cost is financed by debt. The economic and tax-deductible depreciation rates are denoted by $\theta_e$ and $\theta_t$, respectively. Then we have: $\theta_\rho\rho + \theta_t = \mu r$ and $(1-\theta_\rho)\rho + \theta_e - \theta_t = (1-\mu)r \implies \mu = \frac{\theta_\rho\rho+\theta_t}{\rho+\theta_e}$ and $r = \rho + \theta_e$. Following Riedel and Runkel (2007) and Egger et al. (2009), we set: $\rho = 0.05$, $\theta_\rho = 0.4$, $\theta_e = 0.1225$ for machinery, and $\theta_e = 0.0361$ for buildings. The small country's tax rate is assumed to be $t_2^N = 0.10$. From Ernst & Young's survey (see https://www.ey.com/en_gl/technical/tax-guides/worldwide-capital-and-fixed-assets-guide), tax depreciation rates are usually 10-20 percent for machinery and 2-5 percent for buildings. Besides, note that $\frac{r(1-\mu)}{1-t_2^N} = \frac{(1-\theta_\rho)\rho+\theta_e-\theta_t}{1-t_2^N}$, which declines with $\theta_t$. Then setting a small rate $\theta_t = 0.1$ ($\theta_t = 0.02$) for machinery (buildings), we can get: $\frac{\partial k_2^m}{\partial t_m} > 0 \iff \sigma > 5.83\%$ ($\sigma > 5.12\%$). So our analysis suggests that in the short run, the GMT is very likely to raise investments in the small country.

$$
\begin{aligned}
R_2^m &= t_2^N(f_2 - \mu r k_2^m + g^m) + (t_m - t_2^N)(f_2 - \mu r k_2^m + g^m - \sigma k_2^m) \\
&= t_m(f_2 - \mu r k_2^m + g^m) - \underbrace{(t_m - t_2^N)\sigma k_2^m}_{\text{loss from the deduction of SBIE}} .
\end{aligned} \tag{11}
$$

As shown by (11), the introduction of the GMT does not bring country 2's total taxes to the minimum rate $t_m$. The tax amount (i.e., $(t_m - t_2^N)\sigma k_2$ in (6)) saved by affiliate 2 due to the SBIE corresponds to a revenue loss for country 2.

The following proposition presents the revenue effect of the GMT when countries' tax rates are fixed.

**Proposition 1.** *In the short run where countries' tax rates are fixed,*

*(i) the large country benefits from the GMT;*

*(ii) introducing a GMT with minimum rate marginally higher than the small country's equilibrium tax without the GMT increases (reduces) the small country's tax revenue if* $t_2^N > t_2^*$ *(*$t_2^N < t_2^*$*);*

*(iii) (the non-marginal reform) assuming that the small country's revenue function is quasiconcave in* $t_m$*, then the GMT always reduces its tax revenue if* $t_2^N < t_2^*$*,*

*where* $t_2^* := 1 - \sqrt{\frac{r(1-\mu)}{\alpha_2 - \mu r}}$.

*Proof.* See Appendix A.4. □

Intuitively, the introduction of GMT does not change the investment level in country 1 but reduces the outward profit shifting. So the GMT raises the large country's tax revenue. In contrast, the revenue effect of the GMT is ambiguous for the small country. Increasing the minimum rate marginally above the initial tax rate of country 2 affects its tax revenue in the following way:

$$
\left.\frac{\partial R_2^m}{\partial t_m}\right|_{t_m = t_2^N} = \underbrace{\frac{t_2^N(f_2' - \mu r)\sigma}{-(1 - t_2^N)f_2''}}_{\text{gain from tax incentive}} - \underbrace{\sigma k_2^N}_{\text{loss from SBIE}} .
$$

The first term is positive and captures the revenue gain from investment incentives. Recall that due to the SBIE, a higher GMT rate has a tax incentive effect, which exerts investment incentives for affiliate 2, increases its GloBE income and thus raises country 2's revenue. The second term is negative. All else equal, a higher GMT rate increases the top-up tax rate and incurs a larger revenue loss for country 2 due to the deduction of SBIE (cf. (11)). Moreover, for a higher initial tax rate $t_2^N$, the tax incentive effect arising from the marginal rise of the minimum is stronger, while country 2's capital stock and the revenue loss from the SBIE are smaller. Therefore, the revenue gain generated by the marginal tax reform can dominate if and only if country 2's initial tax rate is above a certain threshold.

In the special case of pure profit tax ($\mu = 1$), the threshold $t_2^* = 1$. Then by the proposition, the marginal tax reform always harms the small country. For a pure profit tax, all costs are deductible and investment in each country is undistorted at the initial equilibrium, with the marginal product of capital equaling the exogenous interest rate. Consequently, a higher GMT rate fails to raise affiliate 2's GloBE income through the investment incentives. The marginal reform only causes a larger revenue loss from the deduction of SBIE and thus reduces the small country's revenue.

Part (iii) further shows the global property of the GMT for any minimum rate on the interval $(t_2^N, t_1^N)$, and is more relevant to practical policies. It states that the GMT generally reduces the short-run revenue of the small country if its initial tax rate is low. In Appendix C, we provide sufficient conditions for the quasiconcavity of $R_2^m(t_m)$. Specifically, if $\sigma \leqslant \frac{r(1-\mu)}{1-t_2^N}$ or if $\frac{r(1-\mu)\left(2+t_2^N\right)}{\left(1-t_2^N\right)\left(2-t_2^N\right)} \leqslant \sigma \leqslant \bar{\sigma}^S(t_1^N)$, then $R_2^m(t_m)$ is quasiconcave in $t_m$ for all $t_m \in (t_2^N, t_1^N)$.

Lastly, the following lemma reveals how the profit shifting cost affects the comparison of $t_2^N$ and $t_2^*$.

**Lemma 2.** *There exists a threshold of the concealment cost parameter $\delta^*$ such that $t_2^N > t_2^*$ ($t_2^N \leqslant t_2^*$) if $\delta > \delta^*$ ($\delta \leqslant \delta^*$).*

*Proof.* See Appendix A.5. □

Unlike country 1, country 2's equilibrium tax rate does not necessarily monotonically increase with the concealment cost (see Remark 1). However, $t_2^*$ must intersect $t_2^N(\delta)$ at the upward-sloping part of $t_2^N(\delta)$ (see Appendix A.5). This property directly leads to the lemma.

It immediately follows from Proposition 1 and Lemma 2 that a marginal reform of the GMT will raise (reduce) the small country's short-run revenue under high (low) profit shifting cost. Moreover, assuming the quasiconcavity of the small country's revenue function, then introducing a GMT with minimum rate lying between two countries' initial tax rates always harms the small country in low concealment cost environments.

# 5 Long-run analysis: tax competition

In the long run, both the MNE and the governments adjust their strategies in response to the GMT. In the first stage, two countries simultaneously choose their tax rates, while taking the minimum and carve-out rate as given and anticipating the MNE's responses to their tax choices. In the second stage, the MNE chooses capital investment in each country and the profit shifting level. In the long-run analysis, we assume that the carve-out rate $\sigma \in (\underline{\sigma}, \bar{\sigma}]$, where $\underline{\sigma} := \frac{(1-\mu t_m)r-\alpha_2(1-t_m)}{t_m}$ and $\bar{\sigma} := \frac{\alpha_2(1-t_m)+r[1+(t_m-2)\mu]}{2-t_m}$. As is shown below, the lower bound on the carve-out rate ensures that the small country can attract some investment by setting a low tax rate.[18] It rules out the case where country 2 becomes a tax haven with $k_2 = 0$, $\forall t_2 \in [0,1]$. In Appendix D, we analyze $\sigma \in (0, \underline{\sigma}]$ (which corresponds to a high minimum together with an overly small carve-out) and show that there exists a continuum of Nash equilibria in this case. The upper bound on the carve-out ensures that the excess profit is positive when each country sets its tax below the GMT rate, i.e., $E_i := \pi_i - \sigma k_i \geqslant 0$, $\forall t_i \in [0, t_m)$, $\forall i \in \{1,2\}$ (see Appendix B.2). It rules out the case where the minimum tax is inconsequential for country $i$ due to negative excess profit.

## 5.1 The equilibrium analysis

We first investigate the situation where one country's tax rate is below the GMT rate. Suppose that country $i$'s tax rate $t_i < t_m$. Then the after-tax profit of the MNE reads:

$$\Pi^m = (1-t_m)\left(f_i - \mu r k_i + (-1)^i g\right) - (1-\mu) r k_i + \underbrace{(t_m - t_i)\sigma k_i}_{\text{tax saved from the SBIE}} + (1-t_j)\left(f_j - \mu r k_j + (-1)^j g\right) - (1-\mu) r k_j - \mathbb{1}_j (t_m - t_j)\left(f_j - \mu r k_j + (-1)^j g - \sigma k_j\right) - \frac{\delta}{2} g^2, \tag{12}$$

where indicator variable $\mathbb{1}_j$ equals unity if $t_j < t_m$ and zero otherwise.

[18]For $t_m \leqslant \frac{\alpha_2 - r}{\alpha_2 - \mu r}$, $\sigma > \underline{\sigma}$ is always fulfilled, since in this case $\underline{\sigma} \leqslant 0$.

Solving the profit maximization of the MNE yields:

$$k_i^m = \max\left\{\frac{\alpha_i(1-t_m)-(1-\mu t_m)r+(t_m-t_i)\sigma}{1-t_m}, 0\right\}, \tag{13}$$

$$k_j^m = \begin{cases} \max\left\{\frac{\alpha_j(1-t_j)-(1-\mu t_j)r}{1-t_j}, 0\right\} & \text{if } t_j \geqslant t_m \\ \max\left\{\frac{\alpha_j(1-t_m)-(1-\mu t_m)r+(t_m-t_j)\sigma}{1-t_m}, 0\right\} & \text{if } t_j < t_m \end{cases}, \tag{14}$$

$$g^m = \begin{cases} (-1)^i \min\left\{\frac{t_j-t_m}{\delta}, f_j(k_j^m)-\mu r k_j^m\right\} & \text{if } t_j \geqslant t_m \\ 0 & \text{if } t_j < t_m \end{cases}. \tag{15}$$

From (13), for interior solution with $k_i^m > 0$, we have $\frac{\partial k_i^m}{\partial t_i} < 0$ and $\frac{\partial k_i^m}{\partial \sigma} > 0$. The intuition is straightforward. Due to the deduction of SBIE, affiliate $i$ can reduce its tax payments by $(t_m - t_i)\sigma k_i$, as presented by (12). Hence, either a lower tax rate of country $i$ or a higher carve-out rate increases the tax incentives and raises the capital affiliate $i$ employs. In particular, by reducing its tax rate down to zero, country $i$ can gain a maximum level of investment with $\bar{k}_i := \frac{\alpha_i(1-t_m)-(1-\mu t_m)r+t_m\sigma}{1-t_m}$. Under the assumption $\sigma > \underline{\sigma}$, we have $\bar{k}_2 > 0$, which means that the small country is able to attract inward investment by setting a low corporate tax rate. Combining (14) and (15) indicates that for all $t_i < t_m$, the profit shifting of the MNE is *independent* of country $i$'s tax choice. Whenever country $i$ sets its tax rate below the minimum, the top-up tax is triggered so that affiliate $i$'s GloBE income is always taxed at $t_m$ (cf. (12)). Country $i$ cannot attract more paper profits from country $j$ by further lowering its tax rate. So the GMT places a floor on the taxation of GloBE incomes. It reduces countries' incentives to compete for paper profits and mitigates the corporate tax competition. Moreover, when both countries' tax rates are below the GMT rate, two affiliates' GloBE incomes are taxed at the same rate of $t_m$, so that the true tax differential vanishes. In this case, profit shifting is irrelevant to the MNE's total tax liability and can be eliminated.

For $t_i < t_m$, country $i$'s tax revenue function under the QDMTT is:[19]

$$\begin{aligned} R_i^m(t_i,t_j) &= t_i\left(f_i - \mu r k_i^m + (-1)^i g^m\right) + (t_m - t_i)\left(f_i - \mu r k_i^m + (-1)^i g^m - \sigma k_i^m\right) \\ &= t_m\left(f_i - \mu r k_i^m + (-1)^i g^m\right) - \underbrace{(t_m - t_i)\,\sigma k_i^m}_{\text{loss from the deduction of SBIE}}. \end{aligned} \tag{16}$$

The following lemma presents the effect of one country's tax rate when it is below the minimum rate.

**Lemma 3.** *Suppose that country i's tax rate is below the GMT rate, i.e., $t_i < t_m$. Then country i's tax rate affects its tax revenue in the following way:*

*(i) For $t_m \leqslant t_i^*$, country i's revenue strictly increases with $t_i$ for all $t_i \in [0, t_m)$;*

*(ii) For $t_m > t_i^*$ and $\underline{\sigma} < \sigma \leqslant \sigma_i^m$, country i's revenue decreases with $t_i$ for all $t_i \in [0, t_m)$;*

*(iii) For $t_m > t_i^*$ and $\max\{\underline{\sigma}, \sigma_i^m\} < \sigma \leqslant \bar{\sigma}$, country i's revenue increases (decreases) with $t_i$ when $t_i \in \left[0, \left(1 - \frac{\sigma_i^m}{\sigma}\right) t_m\right)$ ($t_i \in \left(\left(1 - \frac{\sigma_i^m}{\sigma}\right) t_m, t_m\right)$),*

*where $t_i^* := 1 - \sqrt{\frac{r(1-\mu)}{\alpha_i - \mu r}}$ and $\sigma_i^m := \frac{r\left(1 - 2\mu t_m + \mu t_m^2\right) - \alpha_i (1 - t_m)^2}{t_m (2 - t_m)}$.*

*Proof.* See Appendix A.6. □

Part (i) of Lemma 3 implies that one country will never set its tax rate below the minimum when the minimum rate is low. Otherwise, as shown by part (ii) and (iii), it still has incentives to keep a tax rate below the minimum, even possibly at zero. The intuition is as follows. Given (16), starting from the minimum rate, a marginal decrease in one country's tax rate affects its revenue in the following way:

$$-\left.\frac{\partial R_i^m}{\partial t_i}\right|_{t_i = t_m} = \underbrace{-t_m(f_i' - \mu r)\frac{\partial k_i^m}{\partial t_i}}_{>0} - \sigma k_i^m. \tag{17}$$

The first effect in (17) is positive. From (13), a lower tax rate of country $i$ incentivizes affiliate $i$ to employ more capital. This increases affiliate $i$'s GloBE income and tends to raise country $i$'s tax revenue. The second effect in (17) is negative. Reducing country $i$'s tax rate increases the top-up tax rate and thus causes a larger revenue loss for country $i$ due to the deduction of SBIE. Furthermore, a higher $t_m$ renders real investment more sensitive to the tax rate change (i.e., $\frac{\partial}{\partial t_m}\left|\frac{\partial k_i^m}{\partial t_i}\right| = \frac{\sigma}{(1-t_m)^2} > 0$) and so makes the first effect stronger. When $t_m$ is above the threshold rate $t_i^*$, the first effect prevails so that one country can increases its tax revenue by undercutting the minimum. Otherwise, the second effect dominates so that any tax choice below the minimum is a strictly dominated strategy.

[19] For $t_i \geqslant t_m$, the GMT is inactive for affiliate $i$. In this case, country $i$'s revenue function $R_i^m(t_i,t_j)$ takes the same form as (4), where the capital investment $k_i^m$ is given by (2) and the profit shifting level is $g^m = \begin{cases} (-1)^i \min\left\{\frac{t_j - t_i}{\delta}, f_j(k_j^m) - \mu r k_j^m\right\} & \text{if } t_j \geqslant t_i \\ (-1)^j \min\left\{\frac{t_i - \max\{t_m, t_j\}}{\delta}, f_i(k_i^m) - \mu r k_i^m\right\} & \text{if } t_i > t_j \end{cases}.$

Given a minimum with $t_m > t_i^*$, we now analyze the role of the carve-out rate. Suppose that country $i$ sets its tax rate below the minimum. Recall from (13) that either a higher carve-out rate or a lower tax rate promotes the investment. So when the carve-out rate decreases, country $i$ has to reduce its tax rate to offset the negative effect on capital investment in order to maintain the revenue-maximizing capital level with $k_i^m = \frac{\alpha_i - r}{2-t_m}$ (see Appendix A.6). In particular, for a relatively small carve-out rate with $\sigma \leqslant \sigma_i^m$, country $i$ has to reduce its tax rate down to zero to maximize its revenue.

Notably, it directly follows from Lemma 3(ii) and (iii) that: for $t_m > t_i^*$ and $t_j \in [0,1]$, $\arg\max_{t_i \in [0,t_m]} R_i^m(t_i, t_j) = \max\left\{0, \left(1 - \frac{\sigma_i^m}{\sigma}\right) t_m\right\} =: \tilde{t}_i$. Denote by $t_i(t_j)$ the best response function of country $i$ without the GMT,[20] and the long-run equilibrium tax rate of country $i$ under the GMT is denoted by $t_i^m$. We can establish the following:

**Proposition 2.** *After the introduction of the GMT, two countries set equilibrium tax rates in the following way:*

*(i) For $t_m \leqslant t_2^*$, $t_1^m = t_1(t_m)$, $t_2^m = t_m$.*

*(ii) For $t_2^* < t_m \leqslant t_1^*$, $t_1^m = t_1(t_m)$, $t_2^m = \tilde{t}_2$.*

[20] $\forall t_j \in \left[0, \frac{\alpha_1 - r}{\alpha_1 - \mu r}\right]$, $t_i(t_j) \in \left(0, \frac{\alpha_i - r}{\alpha_i - \mu r}\right)$ is the unique solution to $\frac{\partial}{\partial t_i}\left[\varphi_i(t_i) + \frac{t_i(t_j - t_i)}{\delta}\right] = 0$, where $\varphi_i(t_i) := t_i\left[\frac{\alpha_i^2}{2} - \alpha_i \mu r - \frac{r^2(1-\mu t_i)(1-2\mu+\mu t_i)}{2(1-t_i)^2}\right]$ (see Appendix A.1). Strictly speaking, given country $j$'s tax choice $t_j$, $t_i = t_i(t_j)$ is a *candidate* maximum of country $i$'s revenue function $R_i(t_i, t_j)$. From (2), (3) and (4) and using $\frac{\alpha_1 - r}{\alpha_1 - \mu r} \leqslant 2\frac{\alpha_2 - r}{\alpha_2 - \mu r}$, we can show that: $\arg\max_{t_i \in [0,1]} R_i(t_i, t_j) = t_i(t_j)$, if $\frac{t_j - t_i(t_j)}{\delta} \leqslant f_j(k_j) - \mu r k_j$. For $\frac{t_j - t_i(t_j)}{\delta} > f_j(k_j) - \mu r k_j$ (which may occur when $t_j$ is high and so $t_j - t_i(t_j)$ is large, or when $\delta$ is small), the MNE's profit maximization yields a corner solution where all the profit of affiliate $j$ is shifted to country $i$ (i.e., $|g| = f_j(k_j) - \mu r k_j$ by (3)). In this case, choosing $t_i(t_j)$ does not necessarily maximize country $i$'s tax revenue. In Appendix A.1, we rule out the corner solutions and state that the Nash equilibrium without the GMT must be interior.

*(iii)* *For* $t_m > t_1^*$, *two countries' equilibrium taxes are* $(t_1^m = t_1(t_m),\ t_2^m = \tilde{t}_2)$ *when* $R_1(t_1(t_m), t_m) > R_1^m(\tilde{t}_1, \tilde{t}_2)$, *and* $(t_1^m = \tilde{t}_1,\ t_2^m = \tilde{t}_2)$ *when* $R_1(t_1(t_m), t_m) < R_1^m(\tilde{t}_1, \tilde{t}_2)$; *both* $(t_1^m = t_1(t_m),\ t_2^m = \tilde{t}_2)$ *and* $(t_1^m = \tilde{t}_1,\ t_2^m = \tilde{t}_2)$ *are equilibrium tax rates when* $R_1(t_1(t_m), t_m) = R_1^m(\tilde{t}_1, \tilde{t}_2)$.

*(iv)* *A sufficient condition for* $(t_1^m = \tilde{t}_1,\ t_2^m = \tilde{t}_2)$ *to be Nash equilibrium is:* $t_m > t^{**}$ *and* $\max\{\underline{\sigma}, \sigma_1^m\} < \sigma \leqslant \bar{\sigma}$,

*where* $R_1(t_1(t_m), t_m) = \varphi_1(t_1(t_m)) - \frac{t_1(t_m)[t_1(t_m)-t_m]}{\delta}$, $R_1^m(\tilde{t}_1, \tilde{t}_2) = \begin{cases} \frac{(\alpha_1-r)^2}{2(2-t_m)} & \text{if } \sigma > \sigma_1^m \\ \frac{(\alpha_1-r)^2}{2(2-t_m)} - \frac{(\sigma_1^m-\sigma)^2(2-t_m)t_m^2}{2(1-t_m)^2} & \text{if } \sigma \leqslant \sigma_1^m \end{cases}$, *and* $t^{**}$ *is determined by* $\frac{(\alpha_1-r)^2}{2(2-t^{**})} = \bar{R}_1$ *with* $t^{**} \in \left(t_1^*, \bar{t}_1\right)$.

*Proof.* See Appendix A.7. □

**Remark 3.** When $t_m > t_1^*$ and $R_1(t_1(t_m), t_m) = R_1^m(\tilde{t}_1, \tilde{t}_2)$, the two Nash equilibria are Pareto-rankable. Specifically, $(t_1(t_m), \tilde{t}_2)$ Pareto-dominates $(\tilde{t}_1, \tilde{t}_2)$. The two equilibria involve the same investment level in country 2. However, country 2 has additional revenue at the former equilibrium by taxing the profit shifted from country 1, since the two affiliates' GloBE incomes are taxed at different rates. By contrast, the profit shifting is eliminated at the latter equilibrium. So we have: $R_2^m(\tilde{t}_2, t_1(t_m)) > R_2^m(\tilde{t}_2, \tilde{t}_1)$. For simplicity, in what follows we assume that country 1 chooses $t_1^m = t_1(t_m)$ when the two equilibria occur.

**Remark 4.** An immediate consequence of Proposition 2 is that the GMT reduces the MNE's profit shifting at equilibrium. When $t_m \leqslant t_2^*$, the GMT binds country 2 and the tax rate differential is $t_1(t_m) - t_m$. When $t_m > t_2^*$, country 2 undercuts the minimum and triggers the top-up tax so that affiliate 2's GloBE income is still taxed at $t_m$ (cf. (12)). The true tax rate differential is $t_1(t_m) - t_m$ if country 1 sets the tax rate $t_1(t_m)$, and zero if country 1 also undercuts the minimum. Furthermore, $t_1(t_m) - t_m$ strictly decreases with the minimum rate, since $t_1'(t_m) < 1$ by (A.1). So the GMT always narrows the true tax differential at equilibrium and curbs profit shifting.

Our results are substantially different from the traditional minimum taxation literature which treats minimum tax as a lower bound imposed on each country's tax rate. In such settings, the minimum tax always binds the low-tax country, inducing the high-tax country to set tax rate along the unconstrained best response curve (see Kanbur and Keen, 1993; Keen and Konrad, 2013; Hebous and Keen, 2023). The binding minimum rate is also an implicit assumption in Janeba and Schjelderup (2023), who model the GMT as an increase in the tax haven's tax rate. In our paper, for a pure profit tax ($\mu = 1$), the threshold $t_i^* = 1$. Then by Proposition 2(i), the GMT is binding for the small country. For each country, the capital investment is undistorted whenever its pure profit tax rate is above the minimum. Then lowering the tax rate marginally below the minimum is unable to alter the affiliate's GloBE income through investment incentives, but only causes a larger revenue loss (cf. (17)). So neither country has incentives to undercut the minimum. In this special case, the GMT under Pillar Two works in the same way as "a constraint that no tax rate may be set below the minimum level".

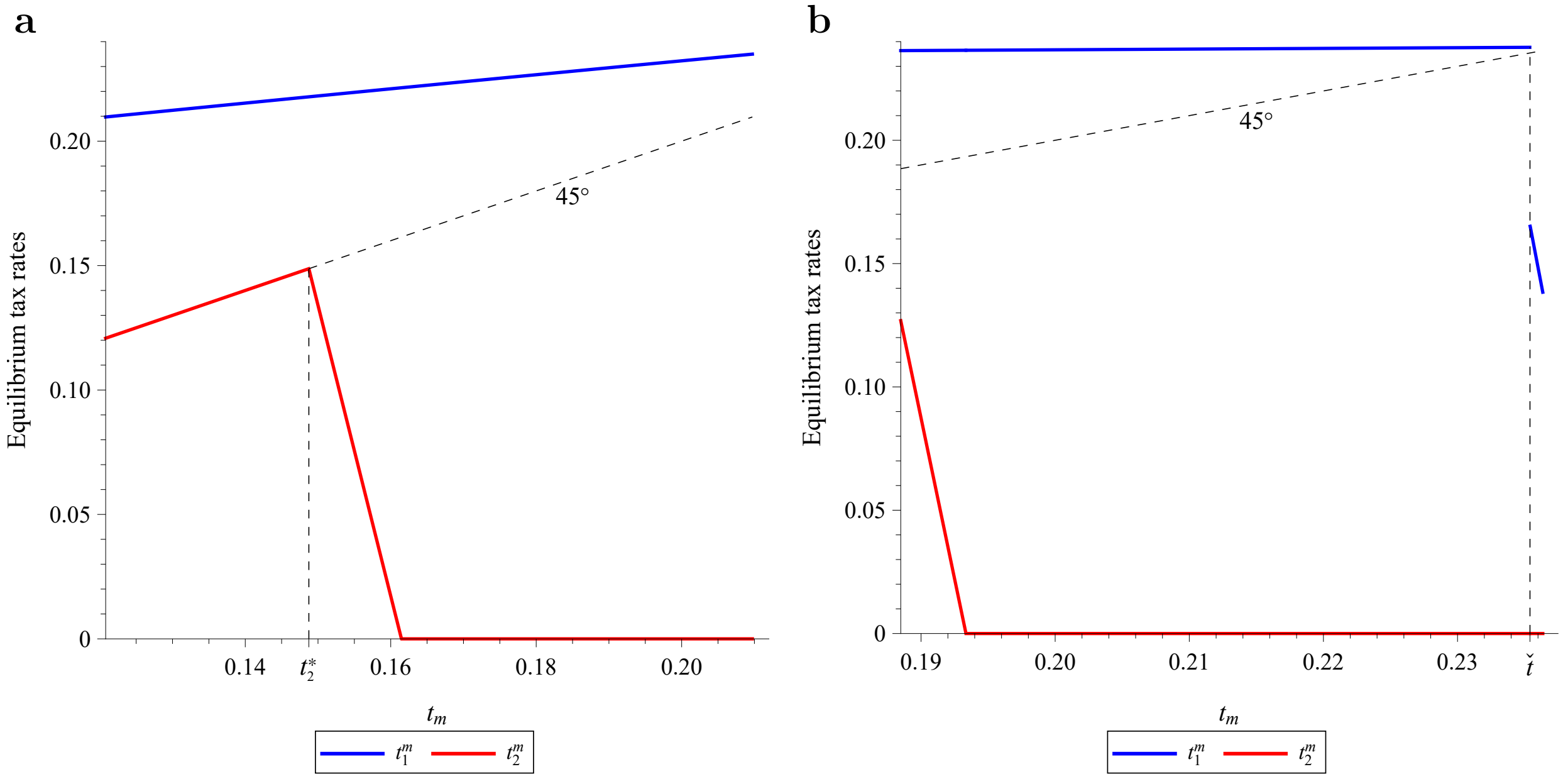

Figure 1: The effect of the minimum rate on equilibrium tax rates

Notes: In panel (a), parameters are specified as: $\alpha_1 = 1.60$, $\alpha_2 = 1.19$, $r = 1$, $\mu = 0.5$, $\delta = 1$, $\sigma = 0.05$. It presents the case where the asymmetry between countries is large and the concealment cost is low. Given the parameter values, we have that: $t_1^N = 0.2097$, $t_2^N = 0.1208$, $t_1^* = 0.3258$, $t_2^* = 0.1487$. The axes origin is defined as $(t_2^N, 0) = (0.1208, 0)$. In panel (b), parameters are specified as: $\alpha_1 = 1.70$, $\alpha_2 = 1.51$, $r = 1$, $\mu = 0$, $\delta = 5$, $\sigma = 0.05$. So compared to panel (a), the market-size asymmetry is smaller and the concealment cost is higher. Given the parameter values, we can obtain that: $t_1^N = 0.2364$, $t_2^N = 0.1885$, $t_1^* = 0.2330$, $t_2^* = 0.1862$, $\check{t} = 0.2354$, where $\check{t}$ is the unique value of $t_m$ satisfying $R_1(t_1(t_m), t_m) = R_1^m(\tilde{t}_1, \tilde{t}_2)$. The axes origin is $(t_2^N, 0) = (0.1885, 0)$.

However, for the general case of partial deductibility ($\mu < 1$), the GMT does not necessarily bind the small country. It is binding only if the minimum rate is low. Otherwise, the small country will set its tax rate below the minimum. Figure 1 illustrates how the long-run equilibrium tax rates vary with the minimum rate, and confirms our analytical results in Proposition 2. As shown in panel (a), country 2 increases its tax rate one-for-one with the minimum rate for $t_m < t_2^* = 0.1487$, while its tax falls below the 45-degree line for $t_m > t_2^*$. In panel (b), country 2 initially sets a tax rate above $t_2^*$ (i.e., $t_2^N > t_2^*$). After introducing the GMT, its equilibrium tax rate is always below the 45-degree line with $t_2^m < t_m$. The intuition is the following. For a minimum rate with $t_m > t_2^*$, country 2 can be better off by undercutting the minimum, since the revenue gain from larger real investments outweighs the revenue loss from the deduction of SBIE (see the discussion below Lemma 3). On the other hand, recall that absent the GMT, country 2's tax base is sensitive to tax rate changes due to the small market size. Consequently, given country 1's equilibrium tax rate, the tax revenue curve of country 2 is downward-sloping for all $t_2 \geqslant t_m$. Starting from the minimum rate, choosing a higher tax rate always reduces country 2's revenue. So when the minimum rate exceeds the threshold $t_2^*$, country 2's best response is to undercut the minimum.[21]

In contrast, the GMT shapes the large country's revenue function in a different way. Since country 1 has a larger market size, its tax base is less sensitive to tax rate changes absent the GMT. Therefore, given country 2's equilibrium tax, country 1's revenue curve is upward-sloping for all $t_1 \in [t_m, t_1(t_m)]$.[22] Starting from the minimum rate, it can be better off by choosing a higher rate $t_1(t_m)$. As illustrated in panel (a), country 1's tax rate moves along the curve $t_1(t_m)$, ranging from $t_1(t_2^N) = 0.2097$ to $t_1(t_1^N) = 0.2350$. In panel (b), with the minimum rate changing from $t_2^N$ to $\check{t}$, country 1's tax increases slightly from 0.2364 to 0.2377. On the other hand, for a high minimum rate with $t_m > t_1^*$, undercutting the minimum is also beneficial for country 1 due to larger inward investments. This sheds new light on the "conventional wisdom" that countries' revenue functions (i.e., the Laffer curves) are unimodal and quasiconcave. In our paper, when $t_m > t_1^*$, country 1's revenue function $R_1^m(t_1, t_2^m)$ becomes *bimodal* with two peaks and is *non-quasiconcave* in $t_1$. In this case, country 1 chooses between undercutting the minimum (with the GloBE income taxed at the minimum rate and more inward investments) and setting a high tax rate $t_1(t_m)$ on the GloBE income (at the cost of outward profit shifting and smaller investment), depending on which tax revenue is larger.[23] In particular, when the GMT rate is sufficiently high and the carve-out is not very small, the benefit of attracting larger investments dominates that of levying a higher tax rate so that country 1 will also undercut the minimum and collect top-up taxes (see Proposition 2(iv)). Panel (b) of Figure 1 presents this case: when $t_m > \check{t} = 0.2354$, the equilibrium tax rate of country 1 also falls below the 45-degree line with $t_1^m < t_m$.[24]

[21] We can further compare the equilibrium investment level before and after the GMT reform. Totally, there are three cases: (i) $t_2^N < t_m \leqslant t_2^*$, (ii) $t_2^* \leqslant t_2^N < t_m$ and (iii) $t_2^N < t_2^* < t_m$. In Case (i), $t_2^m = t_m \implies k_2^m|_{t_2=t_2^m} = \frac{\alpha_2(1-t_m)-(1-\mu t_m)r}{1-t_m} < \frac{\alpha_2\left(1-t_2^N\right)-\left(1-\mu t_2^N\right)r}{1-t_2^N} = k_2^N$. So the investment in country 2 decreases in the long run. For the latter two cases, we have: $t_2^m = \tilde{t}_2$. Assume the carve-out $\sigma > \sigma_2^m$ such that $\tilde{t}_2$ is interior. In Case (ii), $k_2^m|_{t_2=t_2^m} = \frac{\alpha_2-r}{2-t_m} > \frac{\alpha_2-r}{2-t_2^*} = \frac{\alpha_2\left(1-t_2^*\right)-\left(1-\mu t_2^*\right)r}{1-t_2^*} \geqslant k_2^N$. So the investment in country 2 rises in the long run. In Case (iii), the investment effect of the GMT is ambiguous. It is more likely that $k_2^m|_{t_2=t_2^m} > k_2^N$, when $t_2^N$ is closer to $t_2^*$ and $t_m$ is higher.

[22] Given that $t_1'(t_2) > 0$ (see (A.1)), it is straightforward to show that $t_1(t_m) > t_1(t_2^N) = t_1^N > t_m$.

For a given carve-out rate, one country's equilibrium tax is *decreasing* in the GMT rate, whenever it is below the minimum.[25] Recall that a higher GMT rate makes capital investment more sensitive to tax rate changes (i.e., $\frac{\partial}{\partial t_m}\left|\frac{\partial k_i}{\partial t_i}\right| > 0$). This increases the benefit of undercutting the minimum and incentivizes the country to reduce its tax rate, even down to zero (if the carve-out rate $\sigma \leqslant \sigma_i^m$). While country 2's tax rate is continuous in the minimum, gradually increasing the GMT rate may trigger country 1's tax rate to jump *discontinuously* below the minimum rate. In this case, both countries aim at attracting real investment instead of competing for paper profits. These analytical results are confirmed by the numerical simulations. In panel (a) of Figure 1, country 2's tax rate declines with $t_m$ for $t_m > t_2^*$. In particular, it sets a zero tax rate if $t_m \geqslant 0.1615$. In panel (b), country 2's tax rate is decreasing and becomes zero for $t_m \geqslant 0.1933$. Country 1's tax rate suddenly drops when $t_m$ is marginally above $\check{t}$, and it is decreasing in $t_m$ for $t_m > \check{t}$.[26] In summary, the

[23]It follows from (2), (14) and footnote 22 that: $\forall t < t_m$, $k_1^m|_{t_1=t} > k_1^m|_{t_1=t_m} > k_1^m|_{t_1=t_1^N} = k_1^N > k_1^m|_{t_1=t_1(t_m)}$. This indicates that in the long run the GMT reduces (increases) the capital investment in country 1 if $t_1^m = t_1(t_m)$ ($t_1^m = \tilde{t}_1$).

[24]The condition stated in Proposition 2(iv) is sufficient but not necessary. Indeed, for a very small carve-out with $\sigma \leqslant \sigma_1^m$, country 1 may set a zero tax rate after introducing the GMT. For example, suppose that $\sigma = 0.02$, while other parameter values are the same as in panel (b). Then we can still obtain that: for $t_m > \check{t} = 0.2354$, $t_1^m = \tilde{t}_1$. Notably, when $t_m \geqslant 0.2363$, we have: $\sigma = 0.02 \leqslant \sigma_1^m$ and $t_1^m = \tilde{t}_1 = 0$.

[25]If $t_i^m < t_m$, then $t_i^m = \tilde{t}_i = \max\left\{0, \left(1 - \frac{\sigma_i^m}{\sigma}\right) t_m\right\}$ (cf. Proposition 2). As to the interior equilibrium, we can show that: $\frac{\partial t_i^m}{\partial t_m} = \frac{\partial}{\partial t_m}\left[\left(1 - \frac{\sigma_i^m}{\sigma}\right) t_m\right] = \frac{-\alpha_i(3-t_m)(1-t_m)+r\left[(2-t_m)^2\mu-1\right]+\sigma(2-t_m)^2}{(2-t_m)^2\sigma} \leqslant \frac{-\alpha_2(3-t_m)(1-t_m)+r\left[(2-t_m)^2\mu-1\right]+\bar{\sigma}(2-t_m)^2}{(2-t_m)^2\sigma} = -\frac{(\alpha_2-r)(1-t_m)}{(2-t_m)^2\sigma} < 0$.

[26]Formally, we can show that: $\lim_{t_m \to \check{t}^-} t_1^m = t_1(\check{t}) = 0.2377$ and $\lim_{t_m \to \check{t}^+} t_1^m = \tilde{t}_1\big|_{t_m=\check{t}} = 0.1656$.

GMT under Pillar Two will reshape the underlying tax game and the competition pattern between two countries.

## 5.2 The revenue effects of the GMT

In this subsection, we investigate how the GMT affects countries' tax revenues in the long run. First, we show how the comparison of $t_2^N$ and $t_1^*$ is related to the market-size asymmetry and profit shifting cost.

**Lemma 4.** *(i) For $\alpha_2 > \alpha_2^*$, there exists a threshold of the concealment cost parameter $\delta^{**}$ with $\delta^{**} > \delta^*$ such that $t_2^N > t_1^*$ ($t_2^N \leqslant t_1^*$) if $\delta > \delta^{**}$ ($\delta \leqslant \delta^{**}$); (ii) For $\alpha_2 \leqslant \alpha_2^*$, $t_2^N < t_1^*$ always holds, where $\alpha_2^* := \sqrt{(\alpha_1 - \mu r)\left[2\sqrt{r(1-\mu)(\alpha_1 - \mu r)} - r(1-\mu)\right]} + \mu r \in (\underline{\alpha}_2, \alpha_1)$.*

*Proof.* See Appendix A.8. □

Absent the GMT, a reduction in country 2's market size increases the elasticity of its tax base (in absolute value) and thus imposes a downward pressure on its tax rate. When country 2 is sufficiently small (i.e., the market-size asymmetry is sufficiently large), this pressure is so high that country 2's equilibrium tax rate is always below $t_1^*$. Otherwise, $t_1^*$ intersects $t_2^N(\delta)$ at the upward-sloping part of $t_2^N(\delta)$, which directly leads to part (i) of the lemma.

The following proposition reveals how the GMT affects equilibrium tax revenues after the behavioral adjustments of the MNE and two countries in response to the tax reform.

**Proposition 3.** *In the long run where countries' tax rates are endogenously determined,*

*(i) the GMT always increases the large country's tax revenue.*

*(ii) introducing a GMT with minimum rate marginally higher than the small country's equilibrium tax without the GMT increases the small country's tax revenue when $\alpha_2 \in [\underline{\alpha}_2, \alpha_2^*]$, or when $\alpha_2 \in (\alpha_2^*, \alpha_1)$ and $\delta \leqslant \delta^{**}$. Otherwise, it may reduce the small country's tax revenue.*

*(iii) (the non-marginal reform) the GMT can raise the small country's tax revenue whenever the carve-out and minimum rate satisfy $\sigma \in [\sigma_2^m, \bar{\sigma}]$, $t_m \leqslant t_1^*$ and $\epsilon_g(t_m) \in (0, 1]$, where $\epsilon_g(t_m) := -\frac{\partial g^m / g^m}{\partial t_m / t_m}$ is the elasticity of profit shifting with respect to the GMT rate (in absolute value).*

*Proof.* See Appendix A.9. □

**Remark 5.** If the small country loses from a marginal GMT reform in the short run, it can be better off in the long run. This result directly follows from Proposition 1(ii), Lemma 2 and Proposition 3(ii). Introducing the marginal reform under low concealment costs with $\delta < \delta^*$, the small country will experience a transition period of revenue loss. Its tax revenue will be higher than the initial level after countries adjust their tax rates. This implies that it has to take some time until the benefits of the GMT realize for all countries.

The rationale of part (i) is the following. Consider an increase in the GMT rate from $t_2^N$. When $t_m \leqslant t_1^*$, country 1 chooses its tax rate along the initial best-response function. Thus, the revenue effect of the GMT rate for country 1 via the adjustment of its equilibrium tax rate cancels out (by the envelope theorem). On the other hand, a higher GMT rate creates a positive externality on country 1, since it raises the tax rate on affiliate 2' GloBE income and reduces the outward profit shifting in country 1. When $t_m > t_1^*$, country 1 chooses between undercutting the minimum and setting a higher rate $t_1(t_m)$, depending on which tax revenue is larger. In this case, the revenue effect of increasing the minimum rate is no less than that in the case $t_m \leqslant t_1^*$. So introducing the GMT always benefits country 1.

In contrast, the revenue effect for country 2 is more delicate. Part (ii) states that a marginal reform of the GMT increases country 2's long-run revenue either when the market-size asymmetry is *large* or when the profit shifting cost is *low*. The rationale is the following. Consider a marginal increase in the GMT rate from country 2's initial tax rate. For $t_2^N < t_2^*$, the GMT binds country 2 and induces country 1 to set tax rate along the initial best response curve. In this case, our result is the same as the traditional minimum taxation literature which treats the minimum tax as a lower bound imposed on countries' tax rates (see Kanbur and Keen, 1993; Keen and Konrad, 2013; Hebous and Keen, 2023). Absent the GMT country 2's revenue is maximized at $t_2^N$, which means that the marginal reform has a zero first-order effect on its equilibrium revenue (i.e., $\left.\frac{\partial R_2(t_m, t_1(t_m))}{\partial t_2}\right|_{t_m = t_2^N} = 0$). But it induces country 1 to increase tax rate due to the strategic complementarity (i.e., $t_1'(t_m) > 0$ by (A.1)), which in turn imposes a positive externality on country 2's revenue. For $t_2^* \leqslant t_2^N \leqslant t_1^*$, country 2 undercuts the minimum and promotes the real investment to gain a higher tax revenue, while country 1 still chooses its best response $t_1(t_m)$. Consequently, the revenue effect of the marginal reform for country 2 is no less than that in the case $t_2^N < t_2^*$.

Table 1: Simulation results for a marginal GMT reform

| | $t_2^N$ | $R_1(t_1^N, t_2^N)$ | $R_2(t_2^N, t_1^N)$ | $R_1^m(\tilde{t}_1, \tilde{t}_2)\big\|_{t_m=t_2^N}$ | $R_2^m(\tilde{t}_2, \tilde{t}_1)\big\|_{t_m=t_2^N}$ | $(t_1^m, t_2^m)$ |
|---|---|---|---|---|---|---|
| $\delta = 4$ | .2311 | .134593 | .130816 | .134574 | .130702 | $(t_1(t_m), \tilde{t}_2)$ |
| $\delta = 5.5$ | .2334 | .134750 | .130898 | .134754 | .130876 | $(\tilde{t}_1, \tilde{t}_2)$ |
| $\delta = 7$ | .2348 | .134825 | .130929 | .134858 | .130977 | $(\tilde{t}_1, \tilde{t}_2)$ |

Notes: The parameters are fixed as: $\alpha_1 = 1.69$, $\alpha_2 = 1.68$, $r = 1$, $\mu = 0$, $\sigma = .05$. For the specified parameters, we have: $t_1^* = .2308$, $t_2^* = .2285$, $\alpha_2^* = 1.644$ and $\delta^{**} = 3.859$.

In summary, the marginal GMT reform benefits country 2 when $t_2^N \leqslant t_1^*$ (or equivalently, by Lemma 4, when $\alpha_2 \leqslant \alpha_2^*$ or when $\alpha_2 > \alpha_2^*$ and $\delta \leqslant \delta^{**}$).

However, the above argument does not necessarily hold for $t_2^N > t_1^*$ (or equivalently, by Lemma 4, for $\alpha_2 > \alpha_2^*$ and $\delta > \delta^{**}$). As we will show in Table 1, when the asymmetry between two countries is *small* and the profit shifting cost is *intermediate*, even a marginal reform of the GMT under Pillar Two may harm the low-tax country. This is substantially different from the previous minimum taxation literature. To explain the rationale, we decompose country 2's equilibrium revenue as follows:

$$R_2^m(t_2^m, t_1^m) = \underbrace{t_m\left(f_2 - \mu r k_2^m\right) - \left(t_m - t_2^m\right)\sigma k_2^m}_{\text{taxation on true profit}} + \underbrace{t_m g^m(t_m)}_{\text{taxation on shifted profit}}. \tag{18}$$

Table 1 presents a marginal GMT reform for countries with small market-size asymmetry. Row 1 shows the case of a relatively low concealment cost ($\delta = 4$), with $t_2^N > t_1^*$ and $R_1(t_1^N, t_2^N) > R_1^m(\tilde{t}_1, \tilde{t}_2)\big|_{t_m=t_2^N}$. After the marginal reform country 1 sets a tax rate of $t_1(t_m)$, while country 2 undercuts the minimum and gains a higher revenue. By contrast, in rows 2 and 3 we have: $t_2^N > t_1^*$ and $R_1(t_1^N, t_2^N) < R_1^m(\tilde{t}_1, \tilde{t}_2)\big|_{t_m=t_2^N}$. So a marginal increase in the GMT rate from $t_2^N$ induces both countries to set tax rates below the minimum with $(t_1^m, t_2^m) = (\tilde{t}_1, \tilde{t}_2)$. The marginal reform increases the revenue from taxing the true profit generated by substantive activities. However, the true tax rate differential vanishes so that the profit shifted to country 2 jumps *discontinuously* from $\frac{t_1^N - t_2^N}{\delta}$ to zero.[27] This is distinct from the traditional minimum taxation literature in which the minimum rate binds the small country so that the profit shifting level changes continuously with the minimum rate. In row

[27] As shown in Appendix A.9, after introducing the GMT the first term in (18) increases if $\sigma \geqslant \sigma_2^m$, while the second term in (18) reduces to zero if $t_1^m = \tilde{t}_1$.

2, $R_2^m(\tilde{t}_2, \tilde{t}_1)\big|_{t_m=t_2^N} < R_2(t_2^N, t_1^N)$, implying that the marginal GMT reform reduces country 2's revenue. The intuition is clear. When the concealment cost is not very large ($\delta = 5.5$), country 2 initially attracts considerable paper profits. After the marginal tax reform, the revenue loss from eliminating profit shifting dominates the increased taxation on true profits so that country 2's revenue decreases.

Part (iii) provides sufficient conditions for the Pareto-improving non-marginal reform. The intuition is straightforward. For $\sigma \geqslant \sigma_2^m$, the equilibrium investment level is interior such that country 2's revenue from taxing the true profit is larger than that without the GMT. For $t_m \leqslant t_1^*$, the GloBE incomes of affiliate 1 and 2 are taxed at $t_1(t_m)$ and $t_m$, respectively. It rules out the case where both countries undercut the minimum so that profit shifting jumps discretely to zero. When the profit shifting is not sensitive to the minimum rate changes (i.e., $\epsilon_g(t_m) < 1$), the increase in the minimum rate outweighs the decrease in the inward profit shifting so that country 2's revenue from taxing the shifted profit also increases.

# 6 Welfare considerations

So far, we have assumed that countries aim at maximizing corporate tax revenues. In this section, we extend our analysis to the case where countries maximize their welfare. To this end, we modify the government's objective function to take into account the after-tax profits of the MNE (i.e., the private income). Specifically, country $i$'s welfare function without the GMT reads:[28]

$$W_i = \eta\Pi + \gamma R_i, \tag{19}$$

where $\eta \in \left[0, \frac{1}{2}\right]$ represents the share of the MNE owned by residents of country $i$, and $\gamma \geqslant 1$ denotes the marginal cost of public funds (MCPF).

Incorporating the MNE's profits into the government's objective makes the analysis more complicated, and the complete proofs are tedious. So in the following we resort to numerical simulations to illustrate the effects of the GMT, and then explain the rationale of the results. As we will show, due to the SBIE, the trade-off between incentivizing capital investment and incurring tax revenue loss still holds. Besides, the reasoning we provide below can be applied to various parameter specifications.

[28]As in the baseline model, in what follows we use superscript $m$ to denote the situation where the GMT is introduced. Besides, notice that $\eta$ also represents the weight placed on private income in the welfare function. So throughout the paper, the terms "ownership share in the MNE" and "welfare weight of private income" are used interchangeably.

Table 2: Simulation results for welfare maximization

| | Equilibrium without the GMT | | | | | | Equilibrium under the GMT | | | | | |
|---|---|---|---|---|---|---|---|---|---|---|---|---|
| $\eta$ | $t_1^N$ | $t_2^N$ | $R_1^N$ | $R_2^N$ | $W_1^N$ | $W_2^N$ | $t_1^m$ | $t_2^m$ | $R_1^m$ | $R_2^m$ | $W_1^m$ | $W_2^m$ |
| Panel A: low profit shifting cost ($\delta$ = .45) | | | | | | | | | | | | |
| .00 | .199 | .115 | .1109 | .0351 | .1885 | .0596 | .211 | .150 | .1269 | .0359 | .2158 | .0610 |
| .10 | .191 | .108 | .1079 | .0330 | .2066 | .0792 | .206 | .150 | .1268 | .0339 | .2367 | .0788 |
| .20 | .182 | .102 | .1048 | .0309 | .2256 | .1001 | .199 | .150 | .1265 | .0318 | .2580 | .0971 |
| .35 | .170 | .091 | .0995 | .0276 | .2561 | .1339 | .189 | .144 | .1254 | .0284 | .2907 | .1258 |
| .50 | .156 | .081 | .0937 | .0242 | .2891 | .1710 | .177 | .030 | .1234 | .0243 | .3250 | .1566 |
| Panel B: increased profit shifting cost ($\delta$ = .80) | | | | | | | | | | | | |
| .00 | .269 | .147 | .1466 | .0377 | .2492 | .0642 | .270 | .150 | .1475 | .0379 | .2507 | .0644 |
| .10 | .261 | .140 | .1441 | .0362 | .2628 | .0793 | .263 | .150 | .1474 | .0366 | .2679 | .0796 |
| .20 | .252 | .133 | .1414 | .0345 | .2773 | .0955 | .256 | .150 | .1470 | .0352 | .2855 | .0955 |
| .35 | .237 | .122 | .1365 | .0317 | .3007 | .1224 | .244 | .144 | .1458 | .0330 | .3128 | .1209 |
| .50 | .221 | .110 | .1307 | .0285 | .3265 | .1528 | .230 | .030 | .1438 | .0302 | .3416 | .1486 |

Notes: The parameters are fixed as: $\alpha_1 = 1.85$, $\alpha_2 = 1.24$, $r = 1$, $\mu = .45$, $\gamma = 1.7$, $\sigma = .05$, $t_m = .15$. $W_i^N := W_i(t_i^N, t_j^N)$ and $W_i^m := W_i^m(t_i^m, t_j^m)$ $(i, j = 1, 2,\ i \neq j)$ denote the equilibrium welfare levels before and after introducing the GMT, respectively. For simplicity, we assume symmetric ownership share, i.e., the residents of the two countries own the same share of the MNE. Thus, for $\eta = 0$, the MNE is fully owned by a third party, and welfare maximization boils down to tax revenue maximization as analyzed in the previous sections. For $\eta = .5$, the residents of each country own one half of the MNE. When $0 < \eta < .5$, the ownership is distributed among two countries' residents and a third party. On the other hand, Kleven and Kreiner (2006) estimate that the average MCPF of five European countries (Denmark, France, Germany, Italy and the UK) is 1.71. More recently, Dahlby and Ferede (2012) estimate that the MCPF for the Canadian federal government is 1.71. Based on these studies, we assume in the simulations that $\gamma = 1.7$. Our results are qualitatively similar for variations of the parameter values.

Table 2 presents the equilibrium taxes and welfare of two countries before and after the GMT reform. In both panels, two countries' equilibrium tax rates without the GMT are decreasing in ownership share $\eta$. The intuition is clear. From (1), by the envelope theorem, we have $\frac{\partial \Pi}{\partial t_i} = -\pi_i < 0$, implying that each country's tax rate negatively affects the MNE's after-tax profit. For a larger ownership share in the MNE, countries have stronger concerns about private incomes when maximizing their welfare (cf. (19)), which forces them to lower tax rates.

How the ownership share in the MNE affects the equilibrium taxes after the GMT reform is more interesting. As displayed in Table 2, the minimum tax binds the small country when the ownership share is small ($\eta \leqslant 0.2$), while for a large ownership share ($\eta \geqslant 0.35$) the small country undercuts the minimum and triggers the top-up tax. The intuition is as follows. Given (12), (17) and (19), a marginal decrease in country $i$'s tax rate from the GMT rate affects its welfare in the following way:

$$-\left.\frac{\partial W_i^m(t_i,t_j)}{\partial t_i}\right|_{t_i=t_m} = \eta \underbrace{\sigma k_i^m}_{-\frac{\partial \Pi^m}{\partial t_i}} + \gamma \underbrace{\left[-t_m(f_i' - \mu r)\frac{\partial k_i^m}{\partial t_i} - \sigma k_i^m\right]}_{-\frac{\partial R_i^m}{\partial t_i}}. \tag{20}$$

The first term in (20) is positive. By an envelope argument, a lower tax rate raises the MNE's after-tax profit only through the deduction of the SBIE. Specifically, when country 2 reduces its tax rate, affiliate 2 can save more taxes due to the SBIE and thus the total profit of the MNE increases (cf. (12)). The two terms in the squared brackets in (20) capture the trade-off between incentivizing capital investment and incurring tax revenue loss, as shown in the baseline model. Given the parameter specification in Table 2, with the rise of the ownership share, the effect on private income is stronger so that country 2 may undercut the minimum rate.

In both panels of Table 2, introducing the GMT increases the large country's tax rate, i.e., $t_1^m > t_1^N$. The explanation is as follows. Using (20) it is readily verified that for the parameters specified in Table 2, $-\left.\frac{\partial W_1^m(t_1,t_2)}{\partial t_1}\right|_{t_1=t_m} < 0$. This implies that country 1 has no incentive to undercut the minimum rate, i.e., $t_1^m \geqslant t_m$. Hence, given the small country's tax choice $t_2^m \leqslant t_m$, the large country sets its tax along the initial best-response curve with $t_1^m = t_1(t_m)$.[29] On the other hand, absent the GMT two countries' tax rates are strategic

[29] Absent the GMT, country $i$'s best response function $t_i(t_j)$ is implicitly determined by $\frac{\partial W_i(t_i,t_j)}{\partial t_i} = 0$. Given (1), (4), (12) and noticing footnote 19, it is straightforward to show: for $t_1 \geqslant t_m$ and $t_2^m \leqslant t_m$, $\frac{\partial \Pi^m(t_1,t_2^m)}{\partial t_1} = \left.\frac{\partial \Pi(t_1,t_2)}{\partial t_1}\right|_{t_2=t_m} = -\pi_1(t_1,t_m) < 0$, $\frac{\partial R_1^m(t_1,t_2^m)}{\partial t_1} = \left.\frac{\partial R_1(t_1,t_2)}{\partial t_1}\right|_{t_2=t_m}$. Then using (19) and solving the first-order condition of welfare maximization with the GMT, we have that $\frac{\partial W_1^m(t_1^m,t_2^m)}{\partial t_1^m} = \frac{\partial W_1(t_1^m,t_m)}{\partial t_1^m} = 0 \implies t_1^m = t_1(t_m)$.

complements, as presented by the following:

$$\frac{dt_i(t_j)}{dt_j} = -\frac{\frac{\partial^2 W_i}{\partial t_i \partial t_j}}{\frac{\partial^2 W_i}{\partial t_i^2}} = -\frac{\eta \frac{\partial^2 \Pi}{\partial t_i \partial t_j} + \gamma \frac{\partial^2 R_i}{\partial t_i \partial t_j}}{\eta \frac{\partial^2 \Pi}{\partial t_i^2} + \gamma \frac{\partial^2 R_i}{\partial t_i^2}} = \frac{\frac{\gamma - \eta}{\delta}}{(2\gamma - \eta)\left[\frac{(1-\mu^2 r^2}{(1-t_i)^3} + \frac{1}{\delta}\right] + \gamma \frac{3r^2(1-\mu)^2 t_i}{(1-t_i)^4}} \in \left(0, \frac{1}{2}\right). \tag{21}$$

The intuition is clear. When country $j$ raises its tax rate, the profits shifted from country $i$ decrease and the competitive pressure on country $i$ is mitigated. Given that welfare weight of tax revenue is higher than that of private income, country $i$ has incentives to increase its tax, although doing so reduces the MNE's after-tax profits. Using (21), it is obvious that $t_1^m = t_1(t_m) > t_1(t_2^N) = t_1^N$. Besides, given that $\frac{\partial \Pi^m(t_1, t_2^m)}{\partial t_1} < 0$ (see footnote 29) and by the same argument as without the GMT, it is intuitive that the large country's tax rate with the GMT declines with the ownership share. Table 2 also indicates that the GMT reduces the true tax differential and contains profit shifting, as in the baseline model.[30]

The revenue effect of the GMT for both countries is positive for various ownership shares and profit shifting costs, i.e., $R_i^m(t_i^m, t_j^m) > R_i(t_i^N, t_j^N)$, $i, j = 1, 2$, $i \neq j$. We first analyze the large country. From (4) and footnote 19, it is obvious that $R_1^m(t_1^m, t_2^m) = R_1(t_1(t_m), t_m)$. Given that country 1 sets its tax along the best-response curve, country 2's tax rate affects country 1's revenue in the following way:

$$\frac{dR_1(t_1(t_2), t_2)}{dt_2} = \underbrace{\frac{\partial R_1}{\partial t_1} \cdot \frac{dt_1(t_2)}{dt_2}}_{\text{strategic effect}} + \underbrace{\frac{\partial R_1}{\partial t_2}}_{\text{positive externality}} > 0. \tag{22}$$

The first term captures the strategic effect. Due to the strategic complementarity (cf. (21)), an increase in country 2's tax rate induces country 1 to raise its tax. Recall that welfare maximization ($\eta > 0$) takes into account the MNE's after-tax profits. This exerts downward pressure on country 1's tax rate such that $t_1(t_2)$ is on the upward-sloping part of the Laffer curve.[31] So the rise in country 1's tax is revenue increasing. The second term captures the positive externality on country 1: a higher tax rate on affiliate 2's GloBE income reduces the profits shifted from country 1 and tends to increase country 1's revenue. Using (22), we can derive that: $R_1^m(t_1^m, t_2^m) = R_1(t_1(t_m), t_m) > R_1\left(t_1(t_2^N), t_2^N\right) = R_1(t_1^N, t_2^N)$.

[30]Notice that with the GMT two affiliates' GloBE incomes are taxed at $t_1(t_m)$ and $t_m$, respectively. From (21), $t_1(t_2) - t_2$ is strictly decreasing in $t_2$, which leads to $t_1(t_m) - t_m < t_1(t_2^N) - t_2^N = t_1^N - t_2^N$.

[31]Given (19), the first-order condition of welfare maximization reads $\frac{\partial W_1(t_1,t_2)}{\partial t_1} = \eta \frac{\partial \Pi}{\partial t_1} + \gamma \frac{\partial R_1}{\partial t_1} = 0$. Recall that $\frac{\partial \Pi}{\partial t_1} = -\pi_1 < 0$, so we have $\forall \eta > 0$, $\left.\frac{\partial R_1(t_1,t_2)}{\partial t_1}\right|_{t_1 = t_1(t_2)} > 0$. For revenue maximization ($\eta = 0$), country 1's tax revenue is maximized at $t_1(t_2)$, and hence the strategic effect vanishes as an envelope property (see also Section 5.2).

For the small country, we decompose its equilibrium revenue into two parts, as in the baseline model (see (18)). Introducing the GMT can increase the revenue from taxing the true profit generated by genuine economic activities. But the effect on taxing the shifted profit is ambiguous. In panel A where profit shifting cost is low ($\delta = 0.45$), the GMT sharply raises the tax rate on affiliate 2's GloBE income and narrows the true tax differential.[32] This reduces the inward profit shifting in country 2 to a great extent, so that the taxation on shifted profits decreases. For the parameters specified in Table 2, the effect on taxing true profit dominates so that the total revenue of country 2 rises. In contrast, for high profit shifting cost ($\delta = 0.8$), the reduction in true tax differential is small, since country 2's initial tax rate is close to the minimum rate. So the loss of the profits shifted to country 2 is limited, and the revenue from taxing the shifted profits may also increase after the GMT reform.

Now we focus on the welfare levels of two countries. As indicated by Table 2, the GMT always increases the large country's welfare, even when the welfare weight of private income is very high ($\eta = 0.5$). Intuitively, the GMT narrows the true tax differential, reduces outward profit shifting in country 1 and raises the taxation on true profits generated by real economic activities. For the parameters in Table 2, the tax revenue gain is large and dominates the loss of private income due to increased tax burden on the MNE. In the following, we analyze the welfare effect for country 1 in a different way to show how it is related to the *tax base elasticities*. Given country 2's tax rate $t_2^m \leqslant t_m$, using (1), (4), (12), (19) and noticing footnote 19 we have that:

$$W_1^m(t_1^m, t_2^m) = W_1^m\left(t_1(t_m), t_2^m\right) \;\geqslant\; W_1\left(t_1(t_m), t_m\right). \tag{23}$$

where the last inequality becomes an equation if and only if $t_2^m = t_m$.[33]

Suppose that country 1 chooses its tax along the best-response function $t_1(t_2)$. Given

[32]For low profit shifting costs, two countries face intense competition for profits absent the GMT, so that country 2's initial tax rate $t_2^N$ is very low relative to the minimum rate $t_m$. Recall from (21) that $t_1(t_2) - t_2$ strictly decreases with $t_2$. Hence, the true tax differential with the GMT (i.e., $t_1(t_m) - t_m$) is much smaller than without the GMT (i.e., $t_1^N - t_2^N$).

[33]Given (12), by an envelope argument we have: $\forall t_2 < t_m$, $\frac{\partial \Pi^m}{\partial t_2} = -\sigma k_2^m < 0$, implying that $\Pi^m\left(t_1(t_m), t_2\right) > \Pi^m\left(t_1(t_m), t_m\right) = \Pi\left(t_1(t_m), t_m\right)$ if $t_2 < t_m$. Moreover, footnote 19 indicates that $R_1^m\left(t_1(t_m), t_2\right) = R_1\left(t_1(t_m), t_m\right)$ for all $t_2 \leqslant t_m$. Then it is straightforward to show: $W_1^m\left(t_1(t_m), t_2^m\right) > W_1\left(t_1(t_m), t_m\right)$ if $t_2^m < t_m$.

(1), (4), and (19) and by an envelope argument, the effect of country 2's tax on country 1's revenue absent the GMT is: $\frac{dW_1(t_1(t_2),t_2)}{dt_2} = \frac{\partial W_1}{\partial t_2} = -\eta\pi_2 + \gamma t_1 \frac{\partial \pi_1}{\partial t_2}$. Then using the first-order condition of country 1's welfare maximization $\frac{\partial W_1(t_1,t_2)}{\partial t_1} = -\eta\pi_1 + \gamma\left(\pi_1 + t_1\frac{\partial \pi_1}{\partial t_1}\right) = 0$ and noting that $\frac{\partial \pi_1}{\partial t_2} = \frac{\partial \pi_2}{\partial t_1} = \frac{1}{\delta}$, we can derive the following:

$$\frac{dW_1(t_1(t_2),t_2)}{dt_2} = \gamma\pi_2\left(\frac{\partial \pi_2/\pi_2}{\partial t_1/t_1} - \frac{\partial \pi_1/\pi_1}{\partial t_1/t_1} - 1\right). \tag{24}$$

Combining (23) and (24) immediately leads to:[34]

$$W_1^m(t_1^m, t_2^m) > W_1(t_1(t_2^N), t_2^N) = W_1(t_1^N, t_2^N), \text{ if } \frac{\partial \pi_2/\pi_2}{\partial t_1/t_1} - \frac{\partial \pi_1/\pi_1}{\partial t_1/t_1} > 1. \tag{25}$$

From (25), we establish that: introducing the GMT can improve the large country's welfare if the sum of the cross-border tax base elasticity and the own tax base elasticity (measured positively) is larger than unity. We can show that this sufficient condition is satisfied for all parameter values in Table 2. Details of the numerical analysis are available from the authors upon request.

Our results have policy implications for current Pillar Two implementations. In practice, the future of Pillar Two is still in flux, especially for US MNEs navigating the GMT rules. On 28 June 2025, the G7 issued a statement, which proposed a side-by-side solution regarding the interaction of Pillar Two and the US tax system.[35] This side-by-side approach will exempt US parented MNEs from Pillar Two's income inclusion rule (IIR) and undertaxed profits rule (UTPR) in respect of their domestic and foreign profits. Since most large MNEs — especially those in the digital sector — are US companies, the statement brings uncertainty and challenges to the international coordination against profit shifting. However, it is notable that Pillar Two legislation is already enacted in more than 50 jurisdictions. The G7's statement is not binding for the OECD/G20 Inclusive Framework and does not change jurisdictions' existing enacted laws.[36] More importantly, Pillar Two's QDMTT is not affected by the side-by-side approach. Foreign affiliates of American MNEs are still subject to the QDMTT in local jurisdictions. Furthermore, our analysis implies that for large countries the potential gains of implementing the GMT are huge. Indeed, due to the loopholes in current tax systems, the concealment cost $\delta$ is low and thus the two tax base elasticities in (25) are high. We infer that under full implementation of the QDMTT, the GMT would be beneficial for high-tax large countries (which might be the case for the US) from both tax revenue and welfare perspectives.[37]

[34] Notice that (25) also holds for $t_1^m < t_m$. By the definition of Nash equilibrium, given country 2's tax rate $t_2^m$ country 1 chooses between undercutting the minimum and setting a higher rate $t_1(t_m)$, depending on which welfare is larger. So for $t_1^m < t_m$, the first equality in (23) is replaced by $W_1^m(t_1^m, t_2^m) \geqslant W_1^m(t_1(t_m), t_2^m)$, while the remaining results are unchanged.

[35] On 20 January 2025, the Trump administration issued an Executive Order, claiming that the OECD's global minimum tax initiative has no force or effect in the US and withdrawing the participation in the OECD's Pillar Two. Moreover, the Executive Order threatened to impose retaliatory taxes on jurisdictions which were deemed to have tax measures discriminating against American companies (see https://www.whitehouse.gov/presidential-actions/2025/01/the-organization-for-economic-co-operation-and-development-oecd-global-tax-deal-global-tax-deal/). Faced with this situation, the G7 proposed a side-by-side solution to the application of Pillar Two to US parented MNEs, which also allowed for the removal of US retaliatory measures (section 899) in the One Big Beautiful Bill Act. For details of the G7's statement, see https://home.treasury.gov/news/press-releases/sb0181.

In contrast, the GMT does not always improve the small country's welfare. As presented by panel A, with a low profit shifting cost ($\delta = 0.45$) the GMT reduces the small country's welfare except for $\eta = 0$. In panel B where profit shifting cost increases ($\delta = 0.8$), the small country gains from the introduction of the GMT for $\eta \leqslant 0.2$.[38] The intuition is as follows. For a low profit shifting cost, the competition for paper profits is intense absent the GMT, and so the initial tax rates are low. In this case, introducing the GMT sharply increases the tax rate on the GloBE incomes of two affiliates, narrows the true tax differential and significantly reduces the after-tax profits of the MNE. So the loss of private income is very likely to outweigh the tax revenue gain such that country 2's welfare declines. Similarly, a larger ownership share in the MNE decreases the initial tax rates and thus incurs a greater loss of private income after the GMT reform. Moreover, this effect is further amplified when computing the welfare, since ownership share is a multiplier of the private income (see (19)). Hence, introducing the GMT will harm country 2 when the ownership share is large.

Remarkably, our results are different from Janeba and Schjelderup (2023), who state that the GMT may lower countries' tax revenues if shifting profits among countries is costly. By contrast, our results reveal that for various profit shifting cost and ownership shares, both countries' revenues increase after the GMT reform. Furthermore, when we focus on countries' welfare, the scope for the Pareto-improving GMT reform increases for a *higher* concealment cost.[39] As a direct policy implication, this suggests that the previous anti-avoidance measures (e.g., the OECD/G20 BEPS project) — which rendered profit shifting costlier for MNEs — can make the GMT more feasible.

[36]The status of Pillar Two implementation in different jurisdictions is compiled by PwC's survey (see https://www.pwc.com/gx/en/services/tax/pillar-two-readiness/country-tracker.html).

[37]Garcia-Bernardo and Janský (2024) exploit a new methodology and the 2017 OECD CbCR data to investigate the distributional consequences of profit shifting. They find that in the US, the effective tax rates (ETRs) for all affiliates and for the affiliates of foreign MNEs are 35.4% and 15.8%, respectively. Newer and more accurate estimates of the ETRs are needed for the assessment of the GMT reform, and we leave this for future research.

[38]More precisely, in row 3 of panel B the welfare levels of country 2 are $W_2^N = 0.095534$ and $W_2^m = 0.095541$ with $W_2^m > W_2^N$.

# 7 Conclusion

This paper has studied how the global minimum tax under Pillar Two of the OECD/G20 Inclusive Framework affects MNEs' behavior and countries' corporate taxes. To the best of our knowledge, our paper is the first to consider partial tax deductibility, the SBIE and the QDMTT simultaneously in a formal model of tax competition between asymmetric countries. In response to the tax policies, the MNE chooses capital investments and profit shifting to maximize total after-tax profits. Under the QDMTT, the country with tax rate below the minimum will collect top-up taxes from the MNE's affiliate operating in its territory. Introducing a GMT with minimum rate lying between two countries' initial tax rates can reduce the true tax differential, curb profit shifting and always raise the large country's tax revenue. The revenue effect for the small country, however, is generally ambiguous. In the short run where countries' tax rates are fixed, due to tax deduction of the SBIE, a higher minimum rate has two opposite effects. On one hand, it means that more taxes can be saved by the low-tax affiliate, which incentivizes the capital investment. On the other hand, it raises the top-up tax rate and thus incurs a larger revenue loss for the small country. We show that the former (latter) effect dominates under high (low) profit shifting cost so that the small country's tax revenue increases (decreases).

In the long run where countries can adjust their tax rates, the GMT reshapes the underlying tax game and the competition pattern. At the equilibrium, the minimum rate binds the small country only if it is low. Otherwise, the small country will set its tax rate below the minimum, even at zero (if the carve-out rate is relatively small), in order to promote real investments in its territory. When the GMT rate is sufficiently high, both countries undercut the minimum and so profit shifting vanishes. In this case, countries aim at attracting capital investments instead of competing for paper profits. Moreover, we show that countries' equilibrium tax rates are decreasing in the minimum rate whenever they are below the minimum. While the equilibrium tax rate of the small country changes continuously with the minimum, gradual increases in the GMT rate may trigger the large country's tax rate to drop discontinuously below the minimum. As for the long-run revenue effect, a marginal GMT reform does not necessarily benefit the small country. It can raise the small country's tax revenue if the market-size asymmetry is large or if the profit shifting cost is low. For small market-size asymmetry and intermediate profit shifting cost, the revenue loss from the elimination of profit shifting may outweigh the revenue gain from taxing the true profits generated by genuine economic activities, so that even a marginal reform may harm the small country. Our results are substantially different from the traditional minimum taxation literature which specifies the minimum tax as a lower bound imposed on countries' tax rates. We also investigate the Pareto-improving non-marginal tax reform. When the carve-out rate is not very small, the GMT rate is not very high and profit shifting is not sensitive to the minimum rate, the GMT can benefit the small country by increasing the taxation of both true profit and shifted profit.

[39] As displayed in Table 2, for a low concealment cost ($\delta = 0.45$) the GMT improves both countries' welfare only when $\eta = 0$. For a higher concealment cost ($\delta = 0.8$), the GMT benefits both countries from a welfare perspective when $\eta \leqslant 0.2$.

In the extended model where countries maximize their welfare, simulation results indicate that: for a moderate GMT rate of 15%, the large country moves along the initial best response curve and sets a higher tax rate after the tax reform. The small country raises its tax to the minimum rate if the welfare weight of private income is low, and undercuts the minimum rate otherwise. Our analysis reveals that high-tax large countries would be main beneficiaries of the GMT, since both their tax revenues and welfare increase after the reform. By contrast, introducing the GMT could reduce small countries' welfare when the welfare weight of private income is high.

Our findings suggest that the introduction of the GMT may not induce all countries to raise their tax rates above the minimum rate of 15%. As some evidence indicates (see footnote 5), small countries may still set their taxes below the minimum to attract inward investments and collect top-taxes. Moreover, introducing the GMT in low concealment cost environments — which might be the current situation due to cross-border tax loopholes — will cause a short period of revenue losses for small countries. But for a moderate GMT rate, their revenues will increase in the long run where countries' tax rates are reset. This implies that it has to take some time until the benefits of the GMT realize for all countries. In the case of welfare-maximizing governments, low-tax small countries are more likely to gain from the GMT reform under higher profit shifting costs. This suggests that the previous anti-avoidance measures (e.g., the OECD/G20 BEPS Project) could make the GMT more feasible.

# Appendix A Proofs and derivations

## A.1 Proof of Lemma 1(i)

Recall from (2) that one country's capital investment is zero when it sets a very high tax rate. From (3), the high-tax affiliate will report zero profit when the tax rate differential is large or when profit shifting cost is very small. The possible corner solutions undermine the smoothness of tax revenue function and complicate the analysis. Our proof will handle this issue and show that the Nash equilibrium must be interior. We proceed in five steps.

**Step 1.** We show that $\pi_i(t_i^N, t_j^N) > 0$, $i = 1, 2$.

Given country $j$'s tax rate, country $i$ can set its tax rate marginally above zero to earn a positive tax revenue. This implies that at equilibrium each country's revenue is positive. So the GloBE income (i.e., taxable profit) of affiliate $i$ is $\pi_i(t_i^N, t_j^N) = f_i(k_i^N) - \mu r k_i^N - \frac{t_i^N - t_j^N}{\delta} > 0$.

**Step 2.** We claim that $t_1^N < \frac{\alpha_1 - r}{\alpha_1 - \mu r}$.

If $t_1^N \geqslant \frac{\alpha_1 - r}{\alpha_1 - \mu r}$, then the capital investment in country 1 is $k_1 = 0$. Besides, in this case no profit is shifted to country 1, irrespective of country 2's tax rate. So country 1's tax base is 0, which contradicts $\pi_1(t_1^N, t_2^N) > 0$. Hence, it must be that $t_1^N < \frac{\alpha_1 - r}{\alpha_1 - \mu r}$.

**Step 3.** We claim that $t_2^N \leqslant \frac{\alpha_2 - r}{\alpha_2 - \mu r}$.

To prove this claim, we analyze the following two cases, respectively. Case (i): $t_1^N \leqslant \frac{\alpha_2 - r}{\alpha_2 - \mu r}$. In this case, $\forall t_2 \geqslant \frac{\alpha_2 - r}{\alpha_2 - \mu r}$, $\pi_2(t_2, t_1^N) = 0 \implies t_2^N < \frac{\alpha_2 - r}{\alpha_2 - \mu r}$. Case (ii): $t_1^N > \frac{\alpha_2 - r}{\alpha_2 - \mu r}$. In this case, $\forall t_2 \geqslant t_1^N$, $\pi_2(t_2, t_1^N) = 0 \implies t_2^N < t_1^N$. On the other hand, if $\frac{\alpha_2 - r}{\alpha_2 - \mu r} < t_2^N < t_1^N$, then country 2's equilibrium revenue is $R_2(t_2^N, t_1^N) = \frac{t_2^N\left(t_1^N - t_2^N\right)}{\delta}$. Using $\frac{\alpha_1 - r}{\alpha_1 - \mu r} \leqslant 2\frac{\alpha_2 - r}{\alpha_2 - \mu r}$ and $t_1^N < \frac{\alpha_1 - r}{\alpha_1 - \mu r}$, we can derive: $\frac{t_1^N}{2} < \frac{\alpha_1 - r}{2(\alpha_1 - \mu r)} \leqslant \frac{\alpha_2 - r}{\alpha_2 - \mu r} \implies \frac{t_2(t_1^N - t_2)}{\delta}$ strictly decreases with $t_2$ for all $t_2 \geqslant \frac{\alpha_2 - r}{\alpha_2 - \mu r}$. Consider a tax rate $t_2'$ that is marginally below $t_2^N$. By the continuity of $\pi_1$ in $t_2$, we get: $\pi_1(t_1^N, t_2') > 0$ and $k_2(t_1') = 0$. Then $R_2(t_2', t_1^N) = \frac{t_2'\left(t_1^N - t_2'\right)}{\delta} > R_2(t_2^N, t_1^N)$, which contradicts the definition of Nash equilibrium. So it must be that $t_2^N \leqslant \frac{\alpha_2 - r}{\alpha_2 - \mu r}$.

**Step 4.** We derive country $i$'s best response to $t_j^N$.

Using (2)–(4), we obtain that: $\forall\, (t_i, t_j) \in A_{ij} := \left\{(t_i, t_j) \in \left[0, \frac{\alpha_i - r}{\alpha_i - \mu r}\right] \times \left[0, \frac{\alpha_j - r}{\alpha_j - \mu r}\right] : \pi_i(t_i, t_j) > 0 \text{ and } \pi_j(t_j, t_i) > 0\right\}$, country $i$'s tax revenue $R_i(t_i, t_j) = \varphi_i(t_i) + \frac{t_i(t_j - t_i)}{\delta}$, where $\varphi_i(t_i) := t_i\left(f_i - \mu r k_i\right) = t_i\left[\frac{\alpha_i^2}{2} - \alpha_i \mu r - \frac{r^2(1-\mu t_i)(1-2\mu+\mu t_i)}{2(1-t_i)^2}\right]$ is the revenue from taxing affiliate $i$'s true profit and $\frac{t_i(t_j - t_i)}{\delta}$ is the revenue gain (loss) from the MNE's profit shifting if $t_j > t_i$ ($t_j < t_i$). From the results shown in Steps 1–3, we have $\left(t_i^N, t_j^N\right) \in A_{ij}$. Besides, it is straightforward to show that: for all $t_j \in \left[0, \frac{\alpha_1 - r}{\alpha_1 - \mu r}\right]$, $\frac{\partial}{\partial t_i}\left[\varphi_i(t_i) + \frac{t_i(t_j - t_i)}{\delta}\right]\Big|_{t_i = 0} = \frac{(\alpha_i - r)(\alpha_i + r - 2\mu r)}{2} + \frac{t_j}{\delta} > 0$, $\frac{\partial}{\partial t_i}\left[\varphi_i(t_i) + \frac{t_i(t_j - t_i)}{\delta}\right]\Big|_{t_i = \frac{\alpha_i - r}{\alpha_i - \mu r}} = -\frac{(\alpha_i - r)(\alpha_i - \mu r)^2}{(1-\mu)r} - \frac{\frac{2(\alpha_i - r)}{\alpha_i - \mu r} - t_j}{\delta} < 0$, and $\frac{\partial^2}{\partial t_i^2}\left[\varphi_i(t_i) + \frac{t_i(t_j - t_i)}{\delta}\right] = -\frac{r^2(1-\mu)^2(2+t_i)}{(1-t_i)^4} - \frac{2}{\delta} < 0$, which leads to the following:

(i) $\forall t_j \in \left[0, \frac{\alpha_1 - r}{\alpha_1 - \mu r}\right]$, $\frac{\partial}{\partial t_i}\left[\varphi_i(t_i) + \frac{t_i(t_j - t_i)}{\delta}\right] = 0$ has a unique solution denoted by $t_i(t_j)$ with $t_i(t_j) \in \left(0, \frac{\alpha_i - r}{\alpha_i - \mu r}\right)$; and (ii) $\varphi_i(t_i) + \frac{t_i(t_j - t_i)}{\delta}$ strictly increases (decreases) with $t_i$ if $t_i < t_i(t_j)$ ($t_i > t_i(t_j)$).

Now we argue by contradiction that $t_i^N = t_i(t_j^N)$. If $t_i^N < t_i(t_j^N)$ ($t_i^N > t_i(t_j^N)$), then consider a tax rate $t_i'$ that is marginally above (below) $t_i^N$. By the continuity of $\pi_i$ and $\pi_j$ in $t_i$, we would get: $\pi_i(t_i', t_j^N) > 0$ and $\pi_j(t_j^N, t_i') > 0$. Together with $t_i' \in \left(0, \frac{\alpha_i - r}{\alpha_i - \mu r}\right)$, it would be that: $\left(t_i', t_j^N\right) \in A_{ij} \implies R_i(t_i', t_j^N) > R_i(t_i^N, t_j^N)$, which contradicts the definition of Nash equilibrium. Therefore, it must be that $t_i^N = t_i(t_j^N)$.

**Step 5.** We establish the existence and uniqueness of the Nash equilibrium.

Given $\frac{\partial}{\partial t_i}\left[\varphi_i(t_i) + \frac{t_i(t_j - t_i)}{\delta}\right] = \varphi_i'(t_i) + \frac{t_j - 2t_i}{\delta} = 0$, using the implicit function theorem yields:

$$\frac{dt_i(t_j)}{dt_j} = \frac{1}{2 - \delta\varphi_i''(t_i)} \in \left(0, \frac{1}{2}\right), \tag{A.1}$$

where $\varphi_i''(t_i) = -\frac{r^2(1-\mu)^2(2+t_i)}{(1-t_i)^4} < 0$.

(A.1) shows that for interior solutions two countries' tax rates are strategic complements. Moreover, it implies that function $T(t_1, t_2) := (t_1(t_2), t_2(t_1))$ is a contraction on the space $\left[0, \frac{\alpha_1 - r}{\alpha_1 - \mu r}\right] \times \left[0, \frac{\alpha_2 - r}{\alpha_2 - \mu r}\right]$. Then by the contraction mapping theorem, there exists a unique fixed point of $T(t_1, t_2)$, that is, a unique Nash equilibrium with $t_i^N = t_i(t_j^N) \in \left(0, \frac{\alpha_i - r}{\alpha_i - \mu r}\right), i = 1, 2$.

**Q.E.D.**

## A.2 The comparative static results

By the continuity of $\pi_i$ and $\pi_j$ , together with $t_i^N \in \left(0, \frac{\alpha_i - r}{\alpha_i - \mu r}\right)$, we can show that: in a small neighborhood of $(t_i^N, t_j^N)$, $(t_i, t_j) \in A_{ij}$ and $R_i(t_i, t_j) = \varphi_i(t_i) + \frac{t_i(t_j - t_i)}{\delta}$. Then totally differentiating $\left.\frac{\partial R_i}{\partial t_i}\right|_{\left(t_i^N, t_j^N\right)} = 0$ yields the comparative statics without the GMT:

$$\frac{\partial t_i^N}{\partial \alpha_i} = \frac{\alpha_i - \mu r}{|J|} \cdot \left[ \frac{r^2 (1-\mu)^2 \left(2 + t_j^N\right)}{\left(1 - t_j^N\right)^4} + \frac{2}{\delta} \right] > 0, \tag{A.2}$$

$$\frac{\partial t_j^N}{\partial \alpha_i} = \frac{\alpha_i - \mu r}{\delta |J|} > 0, \tag{A.3}$$

$$\frac{\partial t_i^N}{\partial \delta} = \frac{1}{|J|} \cdot \left[ \frac{(1-\mu)^2 r^2 (2t_i^N - t_j^N)(2 + t_j^N)}{\delta^2 (1 - t_j^N)^4} + \frac{3t_i^N}{\delta^3} \right], \tag{A.4}$$

where $|J| := \frac{\partial^2 R_1}{\partial t_1^2} \cdot \frac{\partial^2 R_2}{\partial t_2^2} - \frac{\partial^2 R_1}{\partial t_1 \partial t_2} \cdot \frac{\partial^2 R_2}{\partial t_1 \partial t_2} = \prod_{i=1}^{2} \left[ \frac{r^2 (1-\mu)^2 \left(2 + t_i^N\right)}{\left(1 - t_i^N\right)^4} + \frac{2}{\delta} \right] - \frac{1}{\delta^2} > 0$ is the Jacobian determinant of the system $\frac{\partial R_i}{\partial t_i}$ $(i = 1, 2)$ evaluated at $(t_1^N, t_2^N)$. **Q.E.D.**

## A.3 Proof of Lemma 1(ii)

For identical countries with $\alpha_1 = \alpha_2$, it must be that $t_1^N = t_2^N$. On the other hand, using (A.2) and (A.3), we can get: $\frac{\partial (t_1^N - t_2^N)}{\partial \alpha_1} = \frac{\alpha_1 - \mu r}{|J|} \cdot \left[ \frac{r^2 (1-\mu)^2 (2 + t_2)}{(1 - t_2)^4} + \frac{1}{\delta} \right] > 0$, which means that $t_1^N(\alpha_1, \alpha_2) - t_2^N(\alpha_1, \alpha_2)$ is strictly increasing in $\alpha_1$. Hence, we can derive that: $\forall \alpha_1 > \alpha_2$, $t_1^N(\alpha_1, \alpha_2) - t_2^N(\alpha_1, \alpha_2) > t_1^N(\alpha_2, \alpha_2) - t_2^N(\alpha_2, \alpha_2) = 0 \implies \forall \alpha_1 > \alpha_2,\ t_1^N > t_2^N$. **Q.E.D.**

## A.4 Proof of Proposition 1

It follows from (9) that $g^m = \frac{t_1^N - t_m}{\delta} < \frac{t_1^N - t_2^N}{\delta} = g^N$. Together with (4), (7) and (10), we have $R_1^m > R_1(t_1^N, t_2^N)$.

Now we prove (ii). Differentiating (8) with respect to $t_m$ and evaluating it at $t_2^N$ gives:

$$\left.\frac{\partial k_2^m}{\partial t_m}\right|_{t_m = t_2^N} = \frac{\partial k_2^N}{\partial t_2^N} + \underbrace{\frac{\sigma}{-(1 - t_2^N) f_2''}}_{\text{tax incentive effect}}, \tag{A.5}$$

where $\frac{\partial k_2^N}{\partial t_2^N} = \frac{f_2' - \mu r}{(1 - t_2^N) f_2''}$ is the derivative of $k_2$ evaluated at $t_2^N$ in the absence of the GMT.

Differentiating (11) with respect to $t_m$ and using (A.5) yields:

$$\left.\frac{\partial R_2^m}{\partial t_m}\right|_{t_m=t_2^N} = \left.\left[f_2(k_2^m) - \mu r k_2^m + g^m + t_m\left((f_2' - \mu r)\frac{\partial k_2^m}{\partial t_m} + \frac{\partial g^m}{\partial t_m}\right) - \sigma k_2^m - (t_m - t_2^N)\sigma\frac{\partial k_2^m}{\partial t_m}\right]\right|_{t_m=t_2^N}$$
$$= \underbrace{f_2(k_2^N) - \mu r k_2^N + \frac{t_1^N - t_2^N}{\delta} + t_2^N\left((f_2' - \mu r)\frac{\partial k_2^N}{\partial t_2^N} - \frac{1}{\delta}\right)}_{\frac{\partial R_2^N}{\partial t_2^N}=0} + \underbrace{\frac{t_2^N(f_2' - \mu r)\sigma}{-(1-t_2^N)f_2''}}_{\text{gain from tax incentive}} - \underbrace{\sigma k_2^N}_{\text{loss from SBIE}}$$
$$= -\frac{\sigma\left[\alpha_2\left(1-t_2^N\right)^2 - r\left(1 - 2\mu t_2^N + \mu\left(t_2^N\right)^2\right)\right]}{\left(1-t_2^N\right)^2}. \tag{A.6}$$

Using (A.6), it is straightforward to show:

$$\left.\frac{\partial R_2^m}{\partial t_m}\right|_{t_m=t_2^N} \gtreqless 0 \iff t_2^N \gtreqless 1 - \sqrt{\frac{r(1-\mu)}{\alpha_2 - \mu r}} =: t_2^*. \tag{A.7}$$

It immediately follows from (A.7) that: for $t_2^N > t_2^*$ ($t_2^N < t_2^*$), $R_2^m(t_m) > R_2^m(t_2^N) = R_2^N$ ($R_2^m(t_m) < R_2^m(t_2^N) = R_2^N$) when $t_m$ is marginally higher than $t_2^N$.

Lastly, we prove part (iii). Notice that for the continuous function $R_2^m(t_m)$, it is quasiconcave in $t_m$ if and only if one of the following conditions is fulfilled: (i) $R_2^m(t_m)$ is increasing; (ii) $R_2^m(t_m)$ is decreasing; or (iii) there is some $t_m' \in \left(t_2^N, t_1^N\right)$ such that $R_2^m(t_m)$ is increasing for all $t_m \in \left(t_2^N, t_m'\right)$ and is decreasing for all $t_m \in \left(t_m', t_1^N\right)$. On the other hand, recall from (A.7) that $\left.\frac{\partial R_2^m}{\partial t_m}\right|_{t_m=t_2^N} < 0$ if $t_2^N < t_2^*$. Assuming that $R_2^m(t_m)$ is quasiconcave in $t_m$ for all $t_m \in \left(t_2^N, t_1^N\right)$, we can derive that: $R_2^m(t_m)$ must be decreasing in $t_m$ if $t_2^N < t_2^* \implies R_2^m(t_m) < R_2^m(t_2^N) = R_2^N$ holds for all $t_m \in \left(t_2^N, t_1^N\right)$, if $t_2^N < t_2^*$. **Q.E.D.**

## A.5 Proof of Lemma 2

One might hope to prove the lemma by comparing $t_2^N(\delta)$ and $t_2^*$ as $\delta \to 0$ and $\delta \to +\infty$, and then by resorting to the monotonicity of $t_2^N(\delta)$. However, unlike $t_1^N(\delta)$, $t_2^N(\delta)$ is not necessarily increasing in $\delta$ (see footnote 14). To surmount this difficulty, we will invoke the Poincaré-Hopf index theorem, which only requires information on $\frac{dt_2^N(\delta)}{d\delta}$ at $\delta$ that satisfies $t_2^N(\delta) = t_2^*$.

Define $\xi(\delta) := t_2^N(\delta) - t_2^*$. Recall from Appendix A.1 that the equilibrium tax rates have to satisfy $\varphi_i'(t_i^N) - \frac{2t_i^N - t_j^N}{\delta} = 0$ $(i, j \in \{1, 2\}$ and $i \neq j)$. Then it is straightforward to show the following:

$$\lim_{\delta \to 0} t_i^N(\delta) = 0,\ i = 1, 2 \implies \lim_{\delta \to 0} \xi(\delta) < 0. \tag{A.8}$$

$$\varphi_2'(t_2^*) = \frac{\alpha_2 - \mu r}{2} \cdot \left[(\alpha_2 - \mu r) + (1-\mu)\, r - 2\sqrt{(\alpha_2 - \mu r)\,(1-\mu) r}\right] > 0 = \varphi_2'(\lim_{\delta \to +\infty} t_2^N(\delta))$$
$$\implies \lim_{\delta \to +\infty} t_2^N(\delta) > t_2^* \implies \lim_{\delta \to +\infty} \xi(\delta) > 0. \tag{A.9}$$

For all $\delta$ satisfying $t_2^N(\delta) = t_2^*$, using (A.4) we can derive that:

$$\frac{2t_2^N(\delta) - t_1^N(\delta)}{\delta} = \varphi_2'(t_2^N(\delta)) = \varphi_2'(t_2^*) > 0 \implies \frac{dt_2^N(\delta)}{d\delta} > 0 \implies \xi'(\delta) > 0. \tag{A.10}$$

(A.10) indicates that $\xi'(\delta)$ is positive whenever $\xi(\delta) = 0$. Together with (A.8) and (A.9), it follows from the Poincaré-Hopf index theorem that there exists a unique $\delta^*$ such that $\xi(\delta^*) = 0$.[40] Furthermore, given (A.8), (A.9) and the uniqueness of $\delta^*$, it is straightforward to show $t_2^N \gtreqqless t_2^* \iff \delta \gtreqqless \delta^*$. **Q.E.D.**

## A.6 Proof of Lemma 3

Denote $\widehat{t_i} := \min\left\{\frac{\alpha_i(1-t_m)-(1-\mu t_m)r+\sigma t_m}{\sigma}, t_m\right\}$. Given (13), (15) and (16), for $t_i \in [\widehat{t_i}, t_m)$, there is no investment in country $i$, and its revenue is $R_i^m = t_m(-1)^i g^m$, which is independent of $t_i$. For $t_i \in [0, \widehat{t_i})$, inserting (13) and (15) into (16) and differentiating the revenue function with respect to $t_i$ yields:

$$\begin{aligned}\frac{\partial R_i^m}{\partial t_i} &= t_m\left(f_i' - \mu r\right)\frac{\partial k_i^m}{\partial t_i} + \sigma k_i^m - \sigma(t_m - t_i)\frac{\partial k_i^m}{\partial t_i} \\ &= \frac{\sigma\left[\alpha_i\left(1-t_m\right)^2 - r\left(1 - 2\mu t_m + \mu t_m^2\right) + \left(t_m - t_i\right)\left(2 - t_m\right)\sigma\right]}{\left(1-t_m\right)^2}.\end{aligned}$$

[40] We first provide a version of the Poincaré-Hopf index theorem that is often used in game theory. Let $v$: $X \to R^n$, where $X \subset R^n$ is a compact cube, be a continuous function. If all zeros of $v$ are regular (i.e., the Jacobian $|Dv(z)| \neq 0$ at all zeros) and $v$ points outward at the boundary of $X$, then $\sum_{z \in \{z \in X : v(z) = 0\}} \text{sign}\,|Dv(z)| = 1$. As for the problem analyzed here, given (A.8) and (A.9), there must exist $0 < \delta' < \delta''$ such that: (i) $\xi(\delta') < 0$ and $\xi(\delta'') > 0$ (the boundary conditions), and (ii) all zeros of $\xi(\delta)$ are on the interval $[\delta', \delta'']$. Together with (A.10), by the index theorem we have: $\sum_{\delta \in \{\delta :\, \xi(\delta)=0 \text{ and } \delta' \leqslant \delta \leqslant \delta''\}} \text{sign}\,\{\xi'(\delta)\} = 1$. This implies that $\xi(\delta)$ has a unique zero. See Vives (2000) for more details and applications of the index theory.

Moreover, it is readily verified that $\forall t_i \in [0, \widehat{t_i})$, $\frac{\partial^2 R_i^m}{\partial t_i^2} = -\frac{(2-t_m)\sigma^2}{(1-t_m)^2} < 0$. Then it is straightforward to show the following:

(i) For $t_m \leqslant 1 - \sqrt{\frac{(1-\mu)r}{\alpha_i - \mu r}} =: t_i^*$, $\frac{\partial R_i^m}{\partial t_i} > 0$, $\forall t_i \in [0, t_m)$. That is, country $i$'s revenue strictly increases with its tax rate. Besides, we have $R_i^m(t_m, t_j) = \varphi_i(t_m) + t_m(-1)^i g^m$.

(ii) For $t_m > t_i^*$ and $\underline{\sigma} < \sigma \leqslant \frac{r\left(1-2\mu t_m + \mu t_m^2\right) - \alpha_i(1-t_m)^2}{t_m(2-t_m)} =: \sigma_i^m$, $\forall t_i \in [0, \widehat{t_i})$, $\frac{\partial R_i^m}{\partial t_i} < 0$. That is, country $i$'s revenue strictly decreases with $t_i$ when $t_i \in [0, \widehat{t_i})$ and remains unchanged when $t_i \in [\widehat{t_i}, t_m)$. So the revenue-maximizing tax rate is zero, with $R_i^m(0, t_j) = \frac{(\alpha_i - r)^2}{2(2-t_m)} - \frac{(\sigma_i^m - \sigma)^2(2-t_m)t_m^2}{2(1-t_m)^2} + t_m(-1)^i g^m$.

(iii) For $t_m > t_i^*$ and $\max\{\underline{\sigma}, \sigma_i^m\} < \sigma \leqslant \bar{\sigma}$, solving $\frac{\partial R_i^m}{\partial t_i} = 0$ yields $t_i = \left(1 - \frac{\sigma_i^m}{\sigma}\right) t_m \in (0, \widehat{t_i})$. Country $i$'s revenue strictly increases (decreases) with $t_i$ when $t_i \in \left[0, \left(1 - \frac{\sigma_i^m}{\sigma}\right) t_m\right)$ $\left(t_i \in \left(\left(1 - \frac{\sigma_i^m}{\sigma}\right) t_m, \widehat{t_i}\right)\right)$ and remains unchanged when $t_i \in [\widehat{t_i}, t_m)$. So the revenue-maximizing tax rate is $\left(1 - \frac{\sigma_i^m}{\sigma}\right) t_m$, with $k_i^m|_{t_i = \left(1 - \sigma_i^m/\sigma\right)t_m} = \frac{\alpha_i - r}{2 - t_m}$ and $R_i^m((1 - \frac{\sigma_i^m}{\sigma})t_m, t_j) = \frac{(\alpha_i - r)^2}{2(2-t_m)} + t_m(-1)^i g^m$. **Q.E.D.**

## A.7 Proof of Proposition 2

As shown in the proof of Lemma 3, given country $j$'s tax rate, country $i$ can choose $t_i = \min\{\tilde{t}_i, t_m\}$ to earn a positive tax revenue, where $\tilde{t}_i := \max\left\{0, \left(1 - \frac{\sigma_i^m}{\sigma}\right) t_m\right\}$. This implies that each country's equilibrium revenue is positive $\implies$ the GloBE income $\pi_i^m(t_i^m, t_j^m) > 0$ , $i = 1, 2$. Notice that $t_m < t_1^N < \frac{\alpha_1 - r}{\alpha_1 - \mu r}$, which means that the GMT is inactive for affiliate 1 when country 1's tax rate $t_1 \geqslant \frac{\alpha_1 - r}{\alpha_1 - \mu r}$. Then repeating Step 2 in the proof of Lemma 1(i) with superscript $N$ replaced by $m$, we can show $t_1^m < \frac{\alpha_1 - r}{\alpha_1 - \mu r}$. In the following, we prove the four parts of Proposition 2, separately.

### A.7.1 Proof of Proposition 2(i)

Given $t_m \leqslant t_2^* < t_1^*$, it follows from Lemma 3(i) that: for each country, any tax choice below the GMT rate is strictly dominated by $t_m$ such that its equilibrium tax must meet $t_i^m \geqslant t_m$. This means that at equilibrium the GMT is inactive for both affiliates. Each country's equilibrium tax base (GloBE income) is: $\pi_i^m(t_i^m, t_j^m) = f_i(k_i^m) - \mu r k_i^m - \frac{t_i^m - t_j^m}{\delta} > 0$.

**Step 1.** We claim that $t_2^m \leqslant \frac{\alpha_2 - r}{\alpha_2 - \mu r}$.

Notice that $t_2^* < \frac{\alpha_2 - r}{\alpha_2 - \mu r}$. So for $t_m \leqslant t_2^*$, we have $t_m < \frac{\alpha_2 - r}{\alpha_2 - \mu r}$. Moreover, note that $R_i^m(t_i, t_j) = R_i(t_i, t_j)$ when $t_i \geqslant t_m$ and $t_j \geqslant t_m$. Then repeating Step 3 in the proof of Lemma 1(i) with superscript $N$ replaced by $m$, we can get: $t_2^m \leqslant \frac{\alpha_2 - r}{\alpha_2 - \mu r}$.

**Step 2.** We claim that $t_i^m = \max\left\{t_m, t_i(t_j^m)\right\}$.

We analyze two cases, respectively. Case (i): $t_m \geqslant t_i(t_j^m)$. We argue by contradiction that $t_i^m = t_m$. If $t_i^m > t_m \geqslant t_i(t_j^m)$, then consider a tax rate $t_i'$ that is marginally below $t_i^m$. Notice that $\left(t_i^m, t_j^m\right) \in A_{ij}$. Then by the continuity of $\pi_i$ and $\pi_j$ in $t_i$, we have: $\pi_i(t_i', t_j^m) > 0$ and $\pi_j(t_j^m, t_i') > 0 \implies \left(t_i', t_j^m\right) \in A_{ij} \implies R_i^m(t_i', t_j^m) = R_i(t_i', t_j^m) > R_i(t_i^m, t_j^m) = R_i^m(t_i^m, t_j^m)$, which contradicts the definition of Nash equilibrium. So it must be that $t_i^m = t_m$. Case (ii): $t_m < t_i(t_j^m)$. By similar reasoning as in Case (i), we can show $t_i^m = t_i(t_j^m)$. Combining the two cases, we have proved the claim.

**Step 3.** The Nash equilibrium taxes are $t_1^m = t_1(t_m)$, $t_2^m = t_m$.

Using (A.1) and $t_i^m = \max\left\{t_m, t_i(t_j^m)\right\}$, we can derive: $t_1(t_2^m) \geqslant t_1(t_m) > t_1(t_2^N) = t_1^N > t_m \implies t_1^m = t_1(t_2^m)$. Now we argue that $t_2^m \neq t_2(t_1^m)$. If $t_2^m = t_2(t_1^m)$, then together with $t_1^m = t_1(t_2^m)$ we would obtain $t_2^m = t_2^N$, which contradicts $t_2^m \geqslant t_m > t_2^N$. So given $t_2^m = \max\{t_m, t_2(t_1^m)\}$, it must be that $t_2^m = t_m$. So the equilibrium taxes have to satisfy $t_1^m = t_1(t_m)$, $t_2^m = t_m$. On the other hand, it follows from (A.1) that the function $t_2(t_1(x)) - x$ strictly decreases with $x$ for all $x \in \left[0, \frac{\alpha_1 - r}{\alpha_1 - \mu r}\right]$. Then we can derive:

$$\begin{aligned} &t_2(t_1^m) - t_m = t_2(t_1(t_m)) - t_m < t_2(t_1(t_2^N)) - t_2^N = 0 \implies t_2(t_1^m) < t_m \\ &\implies \max\{t_m, t_2(t_1^m)\} = t_m. \end{aligned} \tag{A.11}$$

So country 2's best response to country 1's tax choice $t_1(t_m)$ is choosing $t_m$. Hence, $(t_1(t_m), t_m)$ satisfies the definition of Nash equilibrium. **Q.E.D.**

### A.7.2 Proof of Proposition 2(ii)

Firstly, it follows from Lemma 3(i) that: for country 1, any tax choice below the GMT rate is strictly dominated by $t_m$ such that its equilibrium tax must meet $t_1^m \geqslant t_m$.

**Step 1.** We claim that $t_2^m = \tilde{t}_2$.

We analyze two cases: (i) $t_m \geqslant \frac{\alpha_2 - r}{\alpha_2 - \mu r}$, and (ii) $t_m < \frac{\alpha_2 - r}{\alpha_2 - \mu r}$, separately.

Case (i): $t_m \geqslant \frac{\alpha_2 - r}{\alpha_2 - \mu r}$. In this case, we prove by contradiction that $t_2^m \leqslant t_m$. If $t_2^m > t_m$ and $t_1^m \leqslant t_2^m$, then we would have $\pi_2(t_2^m, t_1^m) = 0$, which contradicts $\pi_2^m(t_2^m, t_1^m) > 0$. If $t_2^m > t_m$ and $t_1^m > t_2^m$, then country 2's equilibrium revenue would be $R_2^m(t_2^m, t_1^m) = \frac{t_2^m\left(t_1^m - t_2^m\right)}{\delta}$. Then by the same reasoning as in Step 3 in the proof of Lemma 1(i), for a tax rate $t_2'$ that is marginally below $t_2^m$, we would get: $R_2^m(t_2', t_1^m) > R_2^m(t_2^m, t_1^m)$, which contradicts the definition of Nash equilibrium. So we have shown $t_2^m \leqslant t_m$. Then it follows from Lemma 3 that $t_2^m = \tilde{t}_2$.

Case (ii): $t_m < \frac{\alpha_2 - r}{\alpha_2 - \mu r}$. We argue by contradiction that $t_2^m < t_m$. If $t_2^m \geqslant t_m$, then repeating Steps 1–3 in the proof of Proposition 2(i) would lead to $t_1^m = t_1(t_m)$ and $t_2^m = t_m$. However, it follows from Lemma 3 that $R_2^m(\tilde{t}_2, t_1^m) > R_2^m(t_m, t_1^m)$, which contradicts the definition of the Nash equilibrium. So country 2's equilibrium tax must satisfy $t_2^m < t_m$. Then using Lemma 3 again, it must be that $t_2^m = \tilde{t}_2$.

**Step 2.** Given $t_2^m = \tilde{t}_2$, country 1's best response is choosing $t_1(t_m)$.

Given $t_2^m = \tilde{t}_2$, country 1's revenue function is: $R_1^m(t_1, t_2^m) = \begin{cases} R_1(t_1, t_m) & \text{if } t_m \leqslant t_1 < \hat{t}_1 \\ 0 & \text{if } t_1 \geqslant \hat{t}_1 \end{cases}$, where $\hat{t}_1 \in \left(t_1(t_m), \frac{\alpha_1 - r}{\alpha_1 - \mu r}\right)$ is the unique value of $t_1$ such that $\pi_1^m(\hat{t}_1, t_2^m) = \left[f_1(k_1) - \mu r k_1 - \frac{t_1 - t_m}{\delta}\right]\Big|_{t_1 = \hat{t}_1} = 0$ (noticing that the expression for $R_1^m(t_1, t_2^m)$ holds for all $t_2^m \in [0, t_m]$). Besides, using (A.1) and $t_2^N < t_m < t_1^N$ yields $t_1(t_m) > t_1(t_2^N) = t_1^N > t_m$. Hence, $R_1^m(t_1, t_2^m)$ increases (decreases) with $t_1$ if $t_m \leqslant t_1 < t_1(t_m)$ ($t_1 > t_1(t_m)$), such that country 1's best response is setting a tax rate of $t_1(t_m)$.

**Step 3.** We show that $(t_1(t_m), \tilde{t}_2)$ are Nash equilibrium tax rates.

Given the result shown in Step 2, we only need to analyze country 2's best response to $t_1^m = t_1(t_m)$. Given country 1's tax choice $t_1^m = t_1(t_m)$, country 2's revenue function depends on the relationship between $t_m$, $t_1^m$ and $\frac{\alpha_2 - r}{\alpha_2 - \mu r}$ in the following way:

When $\frac{\alpha_2 - r}{\alpha_2 - \mu r} \leqslant t_m < t_1^m$, $R_2^m(t_2, t_1^m) = \begin{cases} \frac{t_2(t_1^m - t_2)}{\delta} & \text{if } t_m \leqslant t_2 \leqslant t_1^m \\ 0 & \text{if } t_2 > t_1^m \end{cases}$; When $t_m < \frac{\alpha_2 - r}{\alpha_2 - \mu r} < t_1^m$, $R_2^m(t_2, t_1^m) = \begin{cases} R_2(t_2, t_1^m) & \text{if } t_m \leqslant t_2 < \frac{\alpha_2 - r}{\alpha_2 - \mu r} \\ \frac{t_2(t_1^m - t_2)}{\delta} & \text{if } \frac{\alpha_2 - r}{\alpha_2 - \mu r} \leqslant t_2 < t_1^m \\ 0 & \text{if } t_2 \geqslant t_1^m \end{cases}$; When $t_m < t_1^m \leqslant \frac{\alpha_2 - r}{\alpha_2 - \mu r}$, $R_2^m(t_2, t_1^m) = \begin{cases} R_2(t_2, t_1^m) & \text{if } t_m \leqslant t_2 < \hat{t}_2 \\ 0 & \text{if } t_2 \geqslant \hat{t}_2 \end{cases}$, where $\hat{t}_2 \in \left[t_1^m, \frac{\alpha_2 - r}{\alpha_2 - \mu r}\right]$ is the unique value of $t_2$ such that $\pi_2^m(\hat{t}_2, t_1^m) = \left[f_2(k_2) - \mu r k_2 - \frac{t_2 - t_1^m}{\delta}\right]\Big|_{t_2=\hat{t}_2} = 0$. Note that $\frac{t_1^m}{2} < \frac{\alpha_1 - r}{2(\alpha_1 - \mu r)} \leqslant \frac{\alpha_2 - r}{\alpha_2 - \mu r}$, and recall from (A.11) that $t_2(t_1^m) < t_m$. It is straightforward to show that $R_2^m(t_2, t_1^m)$ always decreases with $t_2$ for all $t_2 \geqslant t_m$. Together with Lemma 3, country 2's best response is choosing $\tilde{t}_2$.

**Q.E.D.**

### A.7.3 Proof of Proposition 2(iii)

We proceed in two steps.

**Step 1.** We claim that $t_2^m = \tilde{t}_2$.

If country 1's equilibrium tax rate $t_1^m \geqslant t_m$, then by repeating Step 1 in the proof of Proposition 2(ii), it must be that $t_2^m = \tilde{t}_2$.

If country 1's equilibrium tax $t_1^m < t_m$, then it follows from Lemma 3 that $t_1^m = \tilde{t}_1$. In what follows, we analyze two cases. Case (i): $t_m \geqslant \frac{\alpha_2 - r}{\alpha_2 - \mu r}$. It is obvious that $R_2^m(t_2, t_1^m) = 0$ for all $t_2 \geqslant t_m$. Then using Lemma 3, we have $t_2^m = \tilde{t}_2$. Case (ii): $t_m < \frac{\alpha_2 - r}{\alpha_2 - \mu r}$. For $t_2 \geqslant t_m$, country 2's tax revenue function is $R_2^m(t_2, t_1^m) = \begin{cases} R_2(t_2, t_m) & \text{if } t_m \leqslant t_2 < \hat{t}_2 \\ 0 & \text{if } t_2 \geqslant \hat{t}_2 \end{cases}$, where $\hat{t}_2 \in \left(t_m, \frac{\alpha_2 - r}{\alpha_2 - \mu r}\right)$ is the unique value of $t_2$ such that $\pi_2^m(\hat{t}_2, t_1^m) = \left[f_2(k_2) - \mu r k_2 - \frac{t_2 - t_m}{\delta}\right]\Big|_{t_2=\hat{t}_2} = 0$. Besides, it follows from (A.1) and $t_2^N < t_m < t_1^N$ that: $t_2(t_m) < t_2(t_1^N) = t_2^N < t_m$, which implies that $R_2^m(t_2, t_1^m)$ decreases with $t_2$ for all $t_2 \geqslant t_m$. Together with Lemma 3, country 2's best response to $t_1^m = \tilde{t}_1$ is choosing $\tilde{t}_2$.

**Step 2.** We derive country 1's best response to $t_2^m = \tilde{t}_2$.

Using Lemma 3 and repeating Step 2 in the proof of Proposition 2(ii), we can obtain: $\arg\max_{t_1 \in [0, t_m]} R_1^m(t_1, t_2^m) = \tilde{t}_1$ and $\arg\max_{t_1 \in [t_m, 1]} R_1^m(t_1, t_2^m) = t_1(t_m)$. Therefore, country 1's best response is choosing $t_1(t_m)$ when $R_1^m(t_1(t_m), t_2^m) = R_1(t_1(t_m), t_m) > R_1^m(\tilde{t}_1, t_2^m)$, and $\tilde{t}_1$ when $R_1^m(t_1(t_m), t_2^m) = R_1(t_1(t_m), t_m) < R_1^m(\tilde{t}_1, t_2^m)$. It is indifferent between choosing $t_1(t_m)$ and $\tilde{t}_1$ when $R_1^m(t_1(t_m), t_2^m) = R_1(t_1(t_m), t_m) = R_1^m(\tilde{t}_1, t_2^m)$. **Q.E.D.**

### A.7.4 Proof of Proposition 2(iv)

We first prove $t^{**} \in (t_1^*, \bar{t}_1)$. Using $\frac{(\alpha_1-r)^2}{2(2-t^{**})} = \bar{R}_1 = \varphi_1(\bar{t}_1)$, we can show:

$$\frac{(\alpha_1-r)^2}{2(2-\bar{t}_1)} - \frac{(\alpha_1-r)^2}{2(2-t^{**})} = \frac{(\alpha_1-r)^2}{2(2-\bar{t}_1)} - \varphi_1(\bar{t}_1) = \frac{\left[\alpha_1\left(1-\bar{t}_1\right)^2 - r\left(1-2\mu\bar{t}_1+\mu\bar{t}_1^2\right)\right]^2}{2\left(2-\bar{t}_1\right)\left(1-\bar{t}_1\right)^2} > 0$$

$$\implies t^{**} < \bar{t}_1. \tag{A.12}$$

On the other hand, using $\varphi_1'(\bar{t}_1) = 0$ we can obtain:

$$\frac{\alpha_1}{r} = \sqrt{\frac{1+\bar{t}_1}{\left(1-\bar{t}_1\right)^3}}(1-\mu)+\mu. \tag{A.13}$$

Using (A.13), we have:

$$\bar{R}_1 = \varphi_1(\bar{t}_1) = r^2\bar{t}_1\left[\frac{\alpha_1^2}{2r^2} - \mu\frac{\alpha_1}{r} - \frac{\left(1-\mu\bar{t}_1\right)\left(1-2\mu+\mu\bar{t}_1\right)}{2\left(1-\bar{t}_1\right)^2}\right] = \frac{r^2\left(1-\mu\right)^2\bar{t}_1^2}{\left(1-\bar{t}_1\right)^3}.$$

Then solving $\frac{(\alpha_1-r)^2}{2(2-t^{**})} = \bar{R}_1 = \frac{r^2(1-\mu)^2\bar{t}_1^2}{(1-\bar{t}_1)^3}$ for $t^{**}$ and using (A.13), we can get:

$$t^{**} = 2 - \frac{\left(1-\bar{t}_1\right)^3\left(\frac{\alpha_1}{r}-1\right)^2}{2\left(1-\mu\right)^2\bar{t}_1^2} = 2 - \frac{\left(1-\bar{t}_1\right)^3}{2\bar{t}_1^2}\cdot\left(\sqrt{\frac{1+\bar{t}_1}{\left(1-\bar{t}_1\right)^3}}-1\right)^2. \tag{A.14}$$

Define $H(t) := \frac{\alpha_1}{r}\left(1-t\right)^2 - (1-2\mu t+\mu t^2)$. It is readily verified that $H(t) < 0$ if and only if $t \in \left(t_1^*, 1+\sqrt{\frac{r(1-\mu)}{\alpha_1-\mu r}}\right)$. Plugging (A.13) and (A.14) into $H(t)$ yields:

$$H(t^{**}) = -\frac{(1-\mu)\left(1-\sqrt{1-\bar{t}_1^2}\right)^3\left(5+3\sqrt{1-\bar{t}_1^2}-\bar{t}_1^2\right)}{4\bar{t}_1^4} < 0 \implies t^{**} > t_1^*. \tag{A.15}$$

Combining (A.12) and (A.15) leads to $t^{**} \in (t_1^*, \bar{t}_1)$.

Notice that $\forall t_m \in \left(t_2^N, t_1^N\right)$, $\bar{t}_1 > t_1(t_m) > t_m$. Then replacing $t_1^N$ with $t_1(t_m)$ and $t_2^N$ with $t_m$ in (5), we have:

$$R_1(t_1(t_m), t_m) < \bar{R}_1. \tag{A.16}$$

Let $\delta$ be an arbitrary element of set $\left\{\delta > 0 : t_1^N(\delta) > t^{**}\right\}$, which is nonempty according to (A.12). When $\max\left\{t_2^N(\delta), t^{**}\right\} < t_m < t_1^N(\delta)$ and $\max\left\{\underline{\sigma}, \sigma_1^m\right\} < \sigma \leqslant \bar{\sigma}$, by (A.15) and (A.16) we can show that:

$$t_m > t_1^* \text{ and } R_1^m(\tilde{t}_1, \tilde{t}_2) = \frac{(\alpha_1 - r)^2}{2(2 - t_m)} > \frac{(\alpha_1 - r)^2}{2(2 - t^{**})} = \bar{R}_1 > R_1(t_1(t_m), t_m).$$

Then by Proposition 2(iii), the Nash equilibrium rates are $t_1^m = \tilde{t}_1$, $t_2^m = \tilde{t}_2$. **Q.E.D.**

## A.8 Proof of Lemma 4

As in the proof of Lemma 2, we will again appeal to the Poincaré-Hopf index theorem, which only requires information on $\frac{dt_2^N(\delta)}{d\delta}$ at $\delta$ that satisfies $t_2^N(\delta) = t_1^*$. We proceed in two steps.

**Step 1.** We claim that: for all $\delta$ satisfying $t_2^N(\delta) = t_1^*$, $\frac{dt_2^N(\delta)}{d\delta} > 0$.

First, we prove by contradiction that $2t_2^N > t_1^N$ when $t_2^N = t_1^*$. Recall that equilibrium taxes without the GMT are determined by $\varphi_i'(t_i^N) - \frac{2t_i^N - t_j^N}{\delta} = 0$. If $2t_2^N \leqslant t_1^N$, then we would get:

$$\varphi_1'(2t_2^N) \geqslant \varphi_1'(t_1^N) = \frac{2t_1^N - t_2^N}{\delta} > 0 \geqslant \frac{2t_2^N - t_1^N}{\delta} = \varphi_2'(t_2^N), \tag{A.17}$$

where the first inequality holds since $\varphi_1'(t_1)$ strictly decreases with $t_1$ for all $t_1 \in [0, 1)$.

On the other hand, we could derive the following:

$$\begin{aligned}
\varphi_1'(2t_2^N) - \varphi_2'(t_2^N) &= \frac{1}{2}\left(\left(\alpha_1^2 - 2\mu r\alpha_1\right) - \left(\alpha_2^2 - 2\mu r\alpha_2\right) - \frac{r^2 t_2^N (1-\mu)^2 \left[4 - 9t_2^N + \left(t_2^N\right)^2 + \left(6t_2^N\right)^3\right]}{\left(1 - t_2^N\right)^3 \left(1 - 2t_2^N\right)^3}\right) \\
&\leqslant \frac{1}{2}\left(\left(\alpha_1^2 - 2\mu r\alpha_1\right) - \left(\underline{\alpha}_2^2 - 2\mu r\underline{\alpha}_2\right) - \frac{r^2 t_2^N (1-\mu)^2 \left[4 - 9t_2^N + \left(t_2^N\right)^2 + \left(6t_2^N\right)^3\right]}{\left(1 - t_2^N\right)^3 \left(1 - 2t_2^N\right)^3}\right) \\
&= \frac{r^2 (1-\mu)^2 t_2^N \Omega}{2\left(1 - t_2^N\right)^4 \left(1 - 2t_2^N\right)^3 \left[2 - 2t_2^N + \left(t_2^N\right)^2\right]^2} < 0,
\end{aligned}$$

where the inequality in line 2 holds since $\alpha_2^2 - 2\mu r\alpha_2$ strictly increases with $\alpha_2$ for all $\alpha_2 \geqslant \mu r$, the equality in line 3 is obtained by substitution $\alpha_1 = \frac{r\left[1 - 2\mu t_2^N + \mu\left(t_2^N\right)^2\right]}{\left(1 - t_2^N\right)^2}$ into line 2, and $\Omega := 20t_2^N + 28\left(t_2^N\right)^2 - 151\left(t_2^N\right)^3 + 218\left(t_2^N\right)^4 - 167\left(t_2^N\right)^5 + 82\left(t_2^N\right)^6 - 29\left(t_2^N\right)^7 + 6\left(t_2^N\right)^8 - 8 < 0$, $\forall t_2^N \in (0, 1)$.

This contradicts (A.17). So we have proved that $2t_2^N > t_1^N$ when $t_2^N = t_1^*$. Then using

(A.4), the claim directly follows.

**Step 2.** We prove Lemma 4.

Recall that $\varphi_2'(t_2)$ strictly decreases with $t_2$ for all $t_2 \in [0,1)$. Then we can derive the following:

$$t_1^* \gtreqless \lim_{\delta\to+\infty} t_2^N(\delta) \iff \varphi_2'(t_1^*) \lesseqgtr \varphi_2'(\lim_{\delta\to+\infty} t_2^N(\delta)) = 0 \iff \alpha_2 \lesseqgtr \alpha_2^*,$$

where $\alpha_2^* := \sqrt{(\alpha_1 - \mu r)\left[2\sqrt{r(1-\mu)(\alpha_1-\mu r)} - r(1-\mu)\right]} + \mu r \in (\underline{\alpha}_2, \alpha_1)$.

Define $\zeta(\delta) := t_2^N(\delta) - t_1^*$. Then it follows from the claim shown in Step 1 that:

$$\zeta'(\delta) > 0, \text{ whenever } \zeta(\delta) = 0. \tag{A.18}$$

In what follows, we analyze three cases, respectively.

Case (i): $\alpha_2 > \alpha_2^*$. In this case, $\lim_{\delta\to+\infty} t_2^N(\delta) > t_1^*$. Recall that $\lim_{\delta\to0} t_2^N(\delta) = 0 < t_1^*$. Together with (A.18), by the same reasoning as in the proof of Lemma 2, we can show there exists a threshold $\delta^{**}$ such that $t_2^N \gtreqless t_1^* \iff \delta \gtreqless \delta^{**}$. In addition, by Lemma 2, we can derive that: $t_2^N(\delta^{**}) = t_1^* > t_2^* \implies \delta^{**} > \delta^*$.

Case (ii): $\alpha_2 < \alpha_2^*$. We prove by contradiction that $\forall\delta,\ t_2^N < t_1^*$. Assume that there were some $\delta_0 > 0$ such that $t_2^N(\delta_0) \geqslant t_1^*$. Notice that $\lim_{\delta\to+\infty} t_2^N(\delta) < t_1^*$. Then together with (A.18), there would exist some $\delta'$ and $\delta''$ with $\delta_0 \leqslant \delta' < \delta''$ such that $t_2^N(\delta') > t_1^*$ and $t_2^N(\delta'') < t_1^*$. So we would get: $-\zeta(\delta') < 0$ and $-\zeta(\delta'') > 0$ (the boundary conditions). By the Poincaré-Hopf index theorem, we would have: $\sum_{\delta\in\{\delta:\,\zeta(\delta)=0 \text{ and } \delta'\leqslant\delta\leqslant\delta''\}} \text{sign}\{-\zeta'(\delta)\} = 1$, which contradicts (A.18).

Case (iii): $\alpha_2 = \alpha_2^*$. Using the result shown in Case (ii), we get:

$$\forall\delta,\ t_2^N(\delta, \alpha_2^*) = \lim_{\alpha_2\to(\alpha_2^*)^-} t_2^N(\delta,\alpha_2) \leqslant \lim_{\alpha_2\to(\alpha_2^*)^-} t_1^* = t_1^*. \tag{A.19}$$

Moreover, note that $t_2^N(\delta,\alpha_2^*) \neq t_1^*$. Otherwise, using the claim shown in Step 1 we would get: $t_2^N(\delta',\alpha_2^*) > t_1^*$ when $\delta'$ is marginally larger than $\delta$, which contradicts (A.19). So it must be that $\forall\delta,\ t_2^N(\delta,\alpha_2^*) < t_1^*$. **Q.E.D.**

## A.9 Proof of Proposition 3

By Proposition 2, we have: $R_1^m(t_1^m, t_2^m) = R_1(t_1(t_m), t_m)$ when $t_m \leqslant t_1^*$; and $R_1^m(t_1^m, t_2^m) = \max\left\{R_1(t_1(t_m), t_m), R_1^m(\tilde{t}_1, \tilde{t}_2)\right\} \geqslant R_1(t_1(t_m), t_m)$ when $t_m > t_1^*$. Besides, it is straightforward to show that: $\forall t_m \in \left(t_2^N, t_1^N\right)$, $\frac{dR_1(t_1(t_m),t_m)}{dt_m} = \frac{\partial R_1}{\partial t_2} = \frac{t_1(t_m)}{\delta} > 0 \implies R_1(t_1(t_m), t_m) > R_1(t_1(t_2^N), t_2^N) = R_1(t_1^N, t_2^N)$. Hence, the GMT always increases the large country's tax revenue.

Now we prove (ii). Firstly, we analyze the case of $t_2^N = t_1^*$. It is straightforward to show:

$$R_1(t_1(t_2^N), t_2^N) = R_1(t_1^N, t_2^N) > R_1(t_2^N, t_2^N) = \varphi_1(t_2^N) = \frac{(\alpha_1 - r)^2}{2(2 - t_2^N)}, \tag{A.20}$$

where the inequality holds by the definition of the Nash equilibrium and the last equality is obtained by using $t_2^N = t_1^*$.

Using (A.20) and the continuity of $R_1(t_1(t_m), t_m)$ and $\frac{(\alpha_1 - r)^2}{2(2-t_m)}$, together with Proposition 2(iii), we can derive:

$$\begin{aligned} &\left[R_1(t_1(t_m), t_m) - \frac{(\alpha_1 - r)^2}{2(2 - t_m)}\right]\Bigg|_{t_m = t_2^N} > 0 \implies \text{when } t_m \text{ is marginally above } t_2^N, \\ &R_1(t_1(t_m), t_m) > \frac{(\alpha_1 - r)^2}{2(2 - t_m)} \geqslant R_1^m(\tilde{t}_1, \tilde{t}_2) \text{ and thus } (t_1^m, t_2^m) = \left(t_1(t_m), \tilde{t}_2\right). \end{aligned} \tag{A.21}$$

Now consider introducing a GMT with minimum rate $t_m$ marginally above $t_2^N$. It immediately follows from Proposition 2(i), (ii) and (A.21) that: $R_2^m(t_2^m, t_1^m) = R_2(t_m, t_1(t_m))$ if $t_2^N < t_2^*$; and $R_2^m(t_2^m, t_1^m) = R_2^m(\tilde{t}_2, t_1(t_m)) > R_2(t_m, t_1(t_m))$ if $t_2^* \leqslant t_2^N \leqslant t_1^*$. Besides, using (A.1) we can derive: $\frac{dR_2(t_m,t_1(t_m))}{dt_m}\Big|_{t_m=t_2^N} = \left[\frac{\partial R_2}{\partial t_1} \cdot \frac{dt_1(t_m)}{dt_m}\right]\Big|_{t_m=t_2^N} = \frac{t_2^N}{\delta} \cdot \frac{1}{2-\delta\varphi_1''(t_1^N)} > 0 \implies$ when $t_m$ is marginally above $t_2^N$, $R_2(t_m, t_1(t_m)) > R_2(t_2^N, t_1(t_2^N)) = R_2(t_2^N, t_1^N)$. So the marginal reform increases country 2's long-run revenue if $t_2^N \leqslant t_1^*$ (or equivalently, by Lemma 4, if $\alpha_2 \leqslant \alpha_2^*$, or if $\alpha_2 > \alpha_2^*$ and $\delta \leqslant \delta^{**}$).

On the other hand, for the parameter sextuple $(\alpha_1, \alpha_2, \mu, r, \delta, \sigma) = (1.69, 1.68, 0, 1, 5.5, 0.05)$, the second row of Table 1 reveals that: (i) $t_2^N > t_1^*$, (ii) $\left[R_1^m(\tilde{t}_1, \tilde{t}_2) - R_1(t_1(t_m), t_m)\right]\big|_{t_m=t_2^N} = \frac{(\alpha_1-r)^2}{2(2-t_2^N)} - R_1(t_1^N, t_2^N) > 0$, and (iii) $R_2^m(\tilde{t}_2, \tilde{t}_1)\big|_{t_m=t_2^N} - R_2(t_2^N, t_1^N) = \frac{(\alpha_2-r)^2}{2(2-t_2^N)} - R_2(t_2^N, t_1^N) < 0$. Noticing that $R_i^m(\tilde{t}_i, \tilde{t}_j)$ $(i, j = 1, 2;\ i \neq j)$ and $R_1(t_1(t_m), t_m)$ are continuous in $t_m$ and using Proposition 2(iii), we can show the following:[41]

When $t_m$ is marginally above $t_2^N$, (i) $t_m > t_1^*$ and $R_1^m(\tilde{t}_1, \tilde{t}_2) > R_1(t_1(t_m), t_m) \implies t_1^m = \tilde{t}_1$, $t_2^m = \tilde{t}_2$; and (ii) $R_2^m(t_2^m, t_1^m) = R_2^m(\tilde{t}_2, \tilde{t}_1) < R_2(t_2^N, t_1^N)$.

Thus, for the sextuple $(\alpha_1, \alpha_2, \mu, r, \delta, \sigma) = (1.69, 1.68, 0, 1, 5.5, 0.05)$, a marginal GMT reform induces both countries to undercut the minimum and reduces the long-run tax revenue of country 2.

Lastly, we prove (iii). By Proposition 2, country 2's equilibrium tax revenue can be decomposed as follows:

$$R_2^m(t_2^m, t_1^m) = \underbrace{t_m\left(f_2 - \mu r k_2^m\right) - \left(t_m - t_2^m\right)\sigma k_2^m}_{\text{taxation on true profit}} + \underbrace{t_m g^m(t_m)}_{\text{taxation on shifted profit}},$$

where $t_2^m = \begin{cases} t_m & \text{if } t_m \leqslant t_2^* \\ \tilde{t}_2 & \text{if } t_m > t_2^* \end{cases}$, and $g^m(t_m) = \begin{cases} 0 & \text{if } t_m > t_1^* \text{ and } R_1^m(\tilde{t}_1, \tilde{t}_2) > R_1(t_1(t_m), t_m) \\ \frac{t_1(t_m) - t_m}{\delta} & \text{otherwise} \end{cases}$.

For $t_2^N < t_m \leqslant t_2^*$, recall that $\varphi_2'(t_2^*) > 0$ and $\varphi_2''(t_2) < 0$, which leads to: $t_m\left(f_2 - \mu r k_2^m\right) - \left(t_m - t_2^m\right)\sigma k_2^m = \varphi_2(t_m) > \varphi_2(t_2^N)$. For $\sigma \in [\sigma_2^m, \bar{\sigma}]$ and $t_m > \max\left\{t_2^N, t_2^*\right\}$, it is straightforward to show: $t_m\left(f_2 - \mu r k_2^m\right) - \left(t_m - t_2^m\right)\sigma k_2^m = \frac{(\alpha_2 - r)^2}{2(2 - t_m)} > \frac{(\alpha_2 - r)^2}{2(2 - t_2^N)} > \varphi_2(t_2^N)$. Therefore, given that $\sigma \in [\sigma_2^m, \bar{\sigma}]$, the revenue from taxing the true profit is higher than that without the GMT.

On the other hand, when $t_m \leqslant t_1^*$, $\frac{\partial(t_m g^m)}{\partial t_m} = g^m(t_m)\left(1 - \varepsilon_g(t_m)\right)$, where $\varepsilon_g(t_m) := -\frac{\partial g^m / g^m}{\partial t_m / t_m} = \frac{t_m\left(1 - t_1'(t_m)\right)}{t_1(t_m) - t_m} > 0$ is the elasticity of profit shifting with respect to the minimum rate (measured positively). Besides, using (A.1) we can show that: $\frac{d\varepsilon_g(t_m)}{dt_m} = \frac{\left(1 - t_1'(t_m) - t_m t_1''(t_m)\right)\left(t_1(t_m) - t_m\right) + t_m\left(1 - t_1'(t_m)\right)^2}{\left(t_1(t_m) - t_m\right)^2} > 0$, where $t_1''(t_m) = \frac{\delta \varphi_1'''(t_1)}{\left[2 - \delta \varphi_1''(t_1)\right]^3} < 0$ since $\varphi_1'''(t_1) = -\frac{3r^2(3 + t_1)(1 - \mu)^2}{(1 - t_1)^5} < 0$. That is, $\varepsilon_g(t_m)$ strictly increases with $t_m$. Then for any minimum $t_m$ satisfying $t_m \leqslant t_1^*$ and $\varepsilon_g(t_m) \leqslant 1$, by the mean value theorem we have: $t_m g^m - t_2^N g^N = \left(t_m - t_2^N\right) \cdot g^m(\omega)\left[1 - \varepsilon_g(\omega)\right] > 0$, where $\omega \in \left(t_2^N, t_m\right)$. So the revenue from taxing the shifted profit is higher than that without the GMT.

In summary, the conditions for the GMT to increase the small country's revenue are: $\sigma \in [\sigma_2^m, \bar{\sigma}]$, $t_m \leqslant t_1^*$ and $\varepsilon_g(t_m) \in (0, 1]$. **Q.E.D.**

[41] By similar reasoning, it is straightforward to show that in rows 1 and 3, the equilibrium tax rates for the marginal GMT reform are $(t_1(t_m), \tilde{t}_2)$ and $(\tilde{t}_1, \tilde{t}_2)$, respectively.

# Appendix B Conditions for excess profits to be positive

## B.1 The condition for short-run excess profit $E_2 > 0$

In the short run, the excess profit of affiliate 2 is: $E_2 = f_2(k_2^m) - \mu r k_2^m + g^m - \sigma k_2^m$, where $k_2^m$ and $g^m$ are given by (8) and (9), respectively. In what follows, we analyze two cases:

Case (i): $\sigma \leqslant \frac{(1-\mu t_m)r - \alpha_2(1-t_m)}{t_m - t_2^N}$. In this case, there is no investment in country 2, i.e., $k_2^m = 0 \implies E_2 = g^m = \frac{t_1^N - t_m}{\delta} > 0$.

Case (ii): $\sigma > \frac{(1-\mu t_m)r - \alpha_2(1-t_m)}{t_m - t_2^N}$. In this case, $k_2^m = \frac{\alpha_2(1-t_m)-(1-\mu t_m)r+(t_m-t_2^N)\sigma}{1-t_m} > 0 \implies E_2 = f_2(k_2^m) - \mu r k_2^m + g^m - \sigma k_2^m > f_2(k_2^m) - \mu r k_2^m - \sigma k_2^m = k_2^m \cdot \frac{\alpha_2(1-t_m)+r(1-(2-t_m)\mu)-\left(2-t_2^N-t_m\right)\sigma}{2(1-t_m)} \implies E_2 > 0$ if $\sigma \leqslant \frac{\alpha_2(1-t_m)+r(1-(2-t_m)\mu)}{2-t_2^N-t_m} =: \bar{\sigma}^S(t_m)$.

Combining Cases (i) and (ii), a sufficient condition for $E_2 > 0$ is: $\sigma \leqslant \bar{\sigma}^S(t_m)$.

## B.2 The condition for long-run excess profit $E_i \geqslant 0$, $\forall t_i \in [0, t_m)$, $\forall i \in \{1, 2\}$

Suppose that country $i$ sets its tax rate below the minimum, i.e., $t_i < t_m$. Then we have: $E_i = f_i(k_i^m) - \mu r k_i^m + (-1)^i g^m - \sigma k_i^m \geqslant f_i(k_i^m) - \mu r k_i^m - \sigma k_i^m = \left(\alpha_i - \frac{k_i^m}{2} - \mu r - \sigma\right) k_i^m$, where $k_i^m$ and $g_m$ are given by (13) and (15), respectively.

Recall from (13) that $k_i^m$ declines with $t_i$, and note that $k_i^m|_{t_i=0} > 0$. So $\forall t_i \in [0, t_m)$, $E_i \geqslant 0$ if $\alpha_i - \frac{k_i^m}{2}\Big|_{t_i=0} - \mu r - \sigma \geqslant 0$ (or equivalently, if $\sigma \leqslant \frac{\alpha_i(1-t_m)+(1+\mu t_m-2\mu)r}{2-t_m} =: \bar{\sigma}_i$).

Therefore, $\forall t_i \in [0, t_m)$, $\forall i \in \{1, 2\}$, $E_i \geqslant 0$ if and only if $\sigma \leqslant \min\{\bar{\sigma}_1, \bar{\sigma}_2\} = \bar{\sigma}_2 =: \bar{\sigma}$.

# Appendix C Sufficient conditions for the quasiconcavity of $R_2^m(t_m)$, $\forall t_m \in \left(t_2^N, t_1^N\right)$

Firstly, to ensure that the short-run excess profit $E_2 > 0$ for all $t_m \in \left(t_2^N, t_1^N\right)$, we impose an upper bound on the carve-out, i.e., $\sigma \leqslant \inf_{t_m \in \left(t_2^N, t_1^N\right)} \bar{\sigma}^S(t_m) = \bar{\sigma}^S(t_1^N)$. In what follows, we analyze two cases, separately.

Case (a): $\alpha_2(1-t_1^N)-(1-\mu t_1^N)r+(t_1^N-t_2^N)\sigma<0$ (i.e., $\sigma<\frac{(\alpha_2-\mu r)t_1^N-(\alpha_2-r)}{t_1^N-t_2^N}$). Given (8), we distinguish the following two subcases.

Subcase (a1): $t_m\in\left(t_2^N,\frac{\alpha_2-r-\sigma t_2^N}{\alpha_2-\mu r-\sigma}\right)$. We have $k_2^m=\frac{\alpha_2(1-t_m)-(1-\mu t_m)r+(t_m-t_2^N)\sigma}{1-t_m}$. Together with (9) and (11), we can obtain:

$$\frac{d^2R_2^m}{dt_m^2}=-\frac{\left[\left(1-t_2^N\right)\sigma-r\left(1-\mu\right)\right]\left[\left(2-t_2^N\left(4-t_m\right)+t_m\right)\sigma-r\left(2+t_m\right)\left(1-\mu\right)\right]}{\left(1-t_m\right)^4}-\frac{2}{\delta}. \tag{C.1}$$

Besides, it is readily verified that $\frac{(\alpha_2-\mu r)t_1^N-(\alpha_2-r)}{t_1^N-t_2^N}<\frac{r(1-\mu)}{1-t_2^N}<\min\left\{\frac{r(2+t_m)(1-\mu)}{2-t_2^N(4-t_m)+t_m},\bar{\sigma}^S(t_1^N)\right\}$. Then it immediately follows from (C.1) that: $\forall t_m\in\left(t_2^N,\frac{\alpha_2-r-\sigma t_2^N}{\alpha_2-\mu r-\sigma}\right)$, $\frac{d^2R_2^m}{dt_m^2}<0$.

Subcase (a2): $t_m\in\left[\frac{\alpha_2-r-\sigma t_2^N}{\alpha_2-\mu r-\sigma},t_1^N\right)$. We have $k_2^m=0$ and $R_2^m=\frac{t_m\left(t_1^N-t_m\right)}{\delta}$. Besides, we can derive: $\frac{t_1^N}{2}<\frac{\alpha_1-r}{2(\alpha_1-\mu r)}\leqslant\frac{\alpha_2-r}{\alpha_2-\mu r}<\frac{\alpha_2-r-\sigma t_2^N}{\alpha_2-\mu r-\sigma}\implies R_2^m$ strictly decreases with $t_m$ for all $t_m\in\left[\frac{\alpha_2-r-\sigma t_2^N}{\alpha_2-\mu r-\sigma},t_1^N\right)$.

Combining Subcases (a1) and (a2) indicates that: $R_2^m(t_m)$ is quasiconcave for $t_m\in\left(t_2^N,t_1^N\right)$ when $\sigma<\frac{(\alpha_2-\mu r)t_1^N-(\alpha_2-r)}{t_1^N-t_2^N}$.

Case (b): $\alpha_2(1-t_1^N)-(1-\mu t_1^N)r+(t_1^N-t_2^N)\sigma\geqslant 0$ (i.e., $\sigma\geqslant\frac{(\alpha_2-\mu r)t_1^N-(\alpha_2-r)}{t_1^N-t_2^N}$). From (7), for all $t_m\in\left(t_2^N,t_1^N\right)$, $k_2^m=\frac{\alpha_2(1-t_m)-(1-\mu t_m)r+(t_m-t_2^N)\sigma}{1-t_m}$ and $\frac{d^2R_2^m}{dt_m^2}$ is given by (C.1). We distinguish two subcases.

Subcase (b1): $\sigma\in\left[\frac{(\alpha_2-\mu r)t_1^N-(\alpha_2-r)}{t_1^N-t_2^N},\frac{r(1-\mu)}{1-t_2^N}\right]$. Recall that $\frac{r(1-\mu)}{1-t_2^N}<\min\left\{\frac{r(2+t_m)(1-\mu)}{2-t_2^N(4-t_m)+t_m},\bar{\sigma}^S(t_1^N)\right\}$, from which we can show: $\forall t_m\in\left(t_2^N,t_1^N\right)$, $\frac{d^2R_2^m}{dt_m^2}<0$.

Subcase (b2): $\sigma>\frac{r(1-\mu)}{1-t_2^N}$. It is straightforward to show: when $\sigma\geqslant\sup_{t_m\in\left(t_2^N,t_1^N\right)}\frac{r(2+t_m)(1-\mu)}{2-t_2^N(4-t_m)+t_m}=\frac{r(1-\mu)\left(2+t_2^N\right)}{\left(1-t_2^N\right)\left(2-t_2^N\right)}$, $\frac{d^2R_2^m}{dt_m^2}<0$ holds for all $t_m\in\left(t_2^N,t_1^N\right)$.

Combining Subcases (b1) and (b2) indicates that: $R_2^m(t_m)$ is concave (and so quasiconcave) for $t_m\in\left(t_2^N,t_1^N\right)$ when $\sigma\in\left[\frac{(\alpha_2-\mu r)t_1^N-(\alpha_2-r)}{t_1^N-t_2^N},\frac{r(1-\mu)}{1-t_2^N}\right]$, or when $\sigma\in\left[\frac{r(1-\mu)\left(2+t_2^N\right)}{\left(1-t_2^N\right)\left(2-t_2^N\right)},\bar{\sigma}^S(t_1^N)\right]$.

In summary, $R_2^m(t_m)$ is quasiconcave for $t_m\in\left(t_2^N,t_1^N\right)$, if $\sigma\leqslant\frac{r(1-\mu)}{1-t_2^N}$ or if $\frac{r(1-\mu)\left(2+t_2^N\right)}{\left(1-t_2^N\right)\left(2-t_2^N\right)}\leqslant\sigma\leqslant\bar{\sigma}^S(t_1^N)$.

# Appendix D The analysis of the case of $\sigma \in (0, \underline{\sigma}]$

Notice that $0 < \sigma \leqslant \underline{\sigma} \implies t_m > \frac{\alpha_2 - r}{\alpha_2 - \mu r}$. Then it immediately follows from (14) that: $\forall t_2 \in [0,1]$, $k_2^m = 0$. So when $\sigma \leqslant \underline{\sigma}$, introducing the GMT renders country 2 a tax haven where no production takes place and the shifted profits (if any) constitute its only tax base. The following proposition states that there exists a continuum of Nash equilibria in this case.

**Proposition D.1.** *In the case of $\sigma \in (0, \underline{\sigma}]$, after introducing the GMT two countries set equilibrium tax rates in the following way:*

*(i) For $t_m \leqslant t_1^*$, $t_1^m = t_1(t_m)$ and $t_2^m \in [0, t_m]$;*

*(ii) For $t_m > t_1^*$, two countries' equilibrium taxes are $(t_1^m = t_1(t_m), t_2^m \in [0, t_m])$ when $R_1(t_1(t_m), t_m) > R_1^m(\tilde{t}_1, t_2^m)$, $(t_1^m = \tilde{t}_1, t_2^m \in [0, \hat{t}_2^m])$ when $R_1(t_1(t_m), t_m) < R_1^m(\tilde{t}_1, t_2^m) < \bar{R}_1$, and $(t_1^m = \tilde{t}_1,\ t_2^m \in [0,1])$ when $R_1^m(\tilde{t}_1, t_2^m) \geqslant \bar{R}_1$; both $(t_1(t_m), t_2^m \in [0, t_m])$ and $(\tilde{t}_1, t_2^m \in [0, t_m])$ are equilibrium tax rates when $R_1(t_1(t_m), t_m) = R_1^m(\tilde{t}_1, t_2^m)$,*

*where* $\tilde{t}_i := \max\left\{0, \left(1 - \frac{\sigma_i^m}{\sigma}\right) t_m\right\}$, $R_1(t_1(t_m), t_m) = \varphi_1(t_1(t_m)) - \frac{t_1(t_m)[t_1(t_m) - t_m]}{\delta}$,

$$R_1^m(\tilde{t}_1, t_2^m) = \begin{cases} \frac{(\alpha_1 - r)^2}{2(2 - t_m)} & \text{if } \sigma > \sigma_1^m \\ \frac{(\alpha_1 - r)^2}{2(2 - t_m)} - \frac{(\sigma_1^m - \sigma)^2 (2 - t_m) t_m^2}{2(1 - t_m)^2} & \text{if } \sigma \leqslant \sigma_1^m \end{cases}, \text{ and } \hat{t}_2^m \in (t_m, \bar{t}_1).$$

*Proof.* Given country 2's tax rate, country 1 can choose $t_1 = t_m$ to earn a positive revenue, i.e., $R_1^m(t_m, t_2) > 0$. By the same reasoning as in the proof of Proposition 2, we have: $\pi_1^m(t_1^m, t_2^m) > 0$ and $t_1^m < \frac{\alpha_1 - r}{\alpha_1 - \mu r}$.

We first prove part (i). From Lemma 3(i), when $t_m \leqslant t_1^*$, undercutting the minimum is a strictly dominated strategy for country 1. So its equilibrium tax rate must meet $t_1^m \geqslant t_m$. In what follows, we analyze two cases.

Case (i): assume that $0 \leqslant t_2^m \leqslant t_m$. Then repeating Step 2 in the proof of Proposition 2(ii), we have $t_1^m = t_1(t_m)$. On the other hand, given $t_1^m = t_1(t_m)$, country 2's revenue function is: $R_2^m(t_2, t_1^m) = \begin{cases} \frac{t_m(t_1^m - t_m)}{\delta} & \text{if } 0 \leqslant t_2 < t_m \\ \frac{t_2(t_1^m - t_2)}{\delta} & \text{if } t_m \leqslant t_2 < t_1^m \\ 0 & \text{if } t_2 \geqslant t_1^m \end{cases}$. Together with $\frac{t_1^m}{2} < \frac{\alpha_1 - r}{2(\alpha_1 - \mu r)} \leqslant$ $\frac{\alpha_2 - r}{\alpha_2 - \mu r} < t_m$, we can show that any tax rate choice on interval $[0, t_m]$ is country 2's best response.

Case (ii): assume that $t_2^m > t_m$. In this case, notice that: $\forall t_1 \in \left[t_m, \min\left\{\bar{t}_1, t_2^m\right\}\right]$, $R_1^m(t_1, t_2^m) = \varphi_1(t_1)$ with $\varphi_1'(t_1) > 0$. This implies that $t_1^m > t_m$. Recall that the only source of revenues for country 2 is taxing the profit shifted from country 1, which implies that country 2 undercuts country 1, i.e., $t_2^m < t_1^m$. So country 2's equilibrium revenue is $R_2^m(t_2^m, t_1^m) = \frac{t_2^m(t_1^m - t_2^m)}{\delta}$. Then by the same reasoning as in Step 3 in the proof of Lemma 1(i), for a tax rate $t_2'$ that is marginally below $t_2^m$, we get: $R_2^m(t_2', t_1^m) > R_2^m(t_2^m, t_1^m)$, which contradicts the definition of Nash equilibrium. So there is no Nash equilibrium in this case.

Combining Cases (i) and (ii) leads to part (i).

Now we prove part (ii). When $t_m > t_1^*$, we analyze the following three cases.

Case (i): assume that $t_2^m \leqslant t_m$. By the same reasoning as in Step 2 in the proof of Proposition 2(iii), it is straightforward to show: $t_1^m = t_1(t_m)$ when $R_1(t_1(t_m), t_m) > R_1^m(\tilde{t}_1, t_2^m)$, and $t_1^m = \tilde{t}_1$ when $R_1(t_1(t_m), t_m) < R_1^m(\tilde{t}_1, t_2^m)$. Country 1 is indifferent between choosing $t_1(t_m)$ and $\tilde{t}_1$ when $R_1(t_1(t_m), t_m) = R_1^m(\tilde{t}_1, t_2^m)$. On the other hand, for $t_1^m = t_1(t_m)$, country 2's revenue function is the same as in Case (i) in the proof of part (i). For $t_1^m = \tilde{t}_1$, country 2 cannot attract any paper profit from country 1 such that $R_2^m(t_2, t_1^m) = 0, \forall t_2 \in [0, 1]$. Hence, any tax choice on interval $[0, t_m]$ is country 2's best response to $t_1^m$.

Case (ii): assume that $t_2^m > t_m$ and $t_1^m > t_m$. By the same reasoning as in Case (ii) in the proof of Proposition D.1(i), we can show that no Nash equilibrium exists in this case.

Case (iii): assume that $t_2^m > t_m$ and $t_1^m \leqslant t_m$. From Lemma 3, it must be that $t_1^m = \tilde{t}_1$. On the other hand, for $t_1 \in [t_m, 1]$, country 1's revenue function depends on the relationship between $t_2^m$ and $\frac{\alpha_1 - r}{\alpha_1 - \mu r}$ in the following way:

$$
\begin{aligned}
&\text{When } t_2^m \geqslant \frac{\alpha_1 - r}{\alpha_1 - \mu r},\ R_1^m(t_1, t_2^m) = \begin{cases} \varphi_1(t_1) & \text{if } t_m \leqslant t_1 < \frac{\alpha_1 - r}{\alpha_1 - \mu r} \\ 0 & \text{if } t_1 \geqslant \frac{\alpha_1 - r}{\alpha_1 - \mu r} \end{cases}, \\
&\text{When } t_2^m < \frac{\alpha_1 - r}{\alpha_1 - \mu r},\ R_1^m(t_1, t_2^m) = \begin{cases} \varphi_1(t_1) & \text{if } t_m \leqslant t_1 < t_2^m \\ R_1(t_1, t_2^m) & \text{if } t_2^m \leqslant t_1 < \hat{t}_1 \\ 0 & \text{if } t_1 \geqslant \hat{t}_1 \end{cases},
\end{aligned}
\tag{D.1}
$$

where $\hat{t}_1 \in \left(t_2^m, \frac{\alpha_1 - r}{\alpha_1 - \mu r}\right)$ is the unique value of $t_1$ such that $\pi_1^m(\hat{t}_1, t_2^m) = \left[f_1(k_1) - \mu r k_1 - \frac{t_1 - t_2^m}{\delta}\right]\Big|_{t_1 = \hat{t}_1} = 0$.

Given (D.1), solving the maximization problem $\max\limits_{t_1 \in [t_m, 1]} R_1^m(t_1, t_2^m)$ yields:

$$\arg\max_{t_1 \in [t_m, 1]} R_1^m(t_1, t_2^m) = \begin{cases} t_1(t_2^m) & \text{if } t_m < t_2^m \leqslant t_2^{\#} \\ t_2^m & \text{if } t_2^{\#} < t_2^m \leqslant \bar{t}_1 \\ \bar{t}_1 & \text{if } t_2^m > \bar{t}_1 \end{cases}, \tag{D.2}$$

where $t_2^{\#} \in \left(t_1^N, \bar{t}_1\right)$ is the unique value of $t_2$ satisfying $t_1(t_2^{\#}) = t_2^{\#}$.

Plugging (D.2) into (D.1) yields the value function: $V(t_2^m) := \max\limits_{t_1 \in [t_m, 1]} R_1^m(t_1, t_2^m) = \begin{cases} R_1(t_1(t_2^m), t_2^m) & \text{if } t_m < t_2^m \leqslant t_2^{\#} \\ \varphi_1(t_2^m) & \text{if } t_2^{\#} < t_2^m \leqslant \bar{t}_1 \\ \bar{R}_1 & \text{if } t_2^m > \bar{t}_1 \end{cases}$, which is continuous in $t_2^m$ for all $t_2^m \in [t_m, 1]$ and strictly increases with $t_2^m$ for all $t_2^m \in (t_m, \bar{t}_1)$.

Notice that $(\tilde{t}_1, t_2^m)$ are Nash equilibrium tax rates if and only if $R_1^m(\tilde{t}_1, t_2^m) \geqslant V(t_2^m)$, where $R_1^m(\tilde{t}_1, t_2^m) = \begin{cases} \frac{(\alpha_1 - r)^2}{2(2 - t_m)} & \text{if } \sigma > \sigma_1^m \\ \frac{(\alpha_1 - r)^2}{2(2 - t_m)} - \frac{\left(\sigma_1^m - \sigma\right)^2 (2 - t_m) t_m^2}{2(1 - t_m)^2} & \text{if } \sigma \leqslant \sigma_1^m \end{cases}$. Then it is straightforward to show the following:

If $R_1(t_1(t_m), t_m) < R_1^m(\tilde{t}_1, t_2^m) < \bar{R}_1$, then there exists a threshold $\hat{t}_2^m$ such that $t_1^m = \tilde{t}_1$, $t_2^m \in (t_m, \hat{t}_2^m]$ are equilibrium tax rates, where $\hat{t}_2^m \in (t_m, \bar{t}_1)$ is the unique value of $t_2^m$ satisfying $V(\hat{t}_2^m) = R_1^m(\tilde{t}_1, \hat{t}_2^m)$;

If $R_1^m(\tilde{t}_1, t_2^m) \geqslant \bar{R}_1$, then $t_1^m = \tilde{t}_1$, $t_2^m \in (t_m, 1]$ are equilibrium tax rates;

If $R_1^m(\tilde{t}_1, t_2^m) \leqslant R_1(t_1(t_m), t_m)$, then there is no Nash equilibrium.

Lastly, combining Cases (i), (ii) and (iii) leads to part (ii) of the proposition. □

# Appendix E The extension: including labor

In the main text, we have analyzed a simplified version of the GMT where capital is the only (variable) input in the MNE's production process. Now we extend the baseline model to include labor input and provide a complete picture of the GMT. Specifically, the affiliate in country $i$ employs capital $k_i$ and labor $l_i$ to produce a homogenous good according to a general Cobb-Douglas production technology $f_i(k_i, l_i) := k_i^{\lambda} l_i^{\beta}$, with $\lambda, \beta \in (0,1)$ and $\lambda + \beta < 1$. Each affiliate pays a wage rate $w_i$ for each unit of labor. Labor is immobile, and the labor endowment in each country is $\bar{l}_i$ with $\bar{l}_1 > \bar{l}_2 > 0$. With this setup, country 1 (country 2) is the large country (small country) in the sense that country 1 has a larger population. The wage rate is endogenously determined by the labor market clearing condition $l_i = \bar{l}_i$, which equates labor demand $l_i$ and labor endowment $\bar{l}_i$. In line with most countries' corporate tax systems, we assume that payroll cost is fully tax deductible and that a fraction $\mu \in [0, 1)$ of capital cost can be deducted from the corporate tax base. The GloBE income (i.e., taxable profit) of affiliate $i$ is $\pi_i = f_i(k_i, l_i) - \mu r k_i - w_i l_i + (-1)^i g$. The other settings in the baseline model are kept unchanged.

Absent the GMT, the MNE's after-tax profit is:

$$\Pi = \sum_{i=1}^{2} \left[ (1 - t_i) \left( f_i - \mu r k_i - w_i l_i + (-1)^i g \right) - (1 - \mu) \, r k_i \right] - \frac{\delta}{2} g^2.$$

The tax revenue of country $i$ reads:

$$R_i = t_i \pi_i = t_i \left( f_i - \mu r k_i - w_i l_i + (-1)^i g \right).$$

Similar to the baseline model, we can establish the existence and uniqueness of the Nash equilibrium with $t_i^N \in \left(0, \frac{1-\lambda}{1-\mu\lambda}\right)$, $i = 1, 2$. At equilibrium, the small country sets a lower tax rate (i.e., $t_2^N < t_1^N$) to attract paper profits from the large country.[42]

Now consider the introduction of the GMT with rate $t_m \in \left(t_2^N, t_1^N\right)$. In the short run, countries' tax rates are fixed, and only the MNE can adjust its strategies. The GMT is inactive for affiliate 1, while affiliate 2 has to pay a top-up tax, since country 2's tax rate is below the minimum. The SBIE allows the low-tax affiliate to deduct a fraction of the carrying value of tangible assets and payroll expenses from its GloBE income. For simplicity, we assume a common carve-out rate $\sigma$ for capital stock and wage cost. Then the top-up tax paid by affiliate 2 is $(t_m - t_i)[\pi_i - \sigma(k_i + w_i l_i)]$.

[42]The reasoning is similar to that in the baseline model. The proofs and derivations of all results presented in this section are available from the authors upon request.

The MNE's total after-tax profit reads:

$$\Pi^m = \sum_{i=1}^{2}\left[\left(1-t_i^N\right)\underbrace{\left(f_i-\mu rk_i-w_il_i+(-1)^ig\right)}_{\text{GloBE income}}-(1-\mu)\,rk_i\right]$$
$$-\underbrace{\left(t_m-t_2^N\right)\left[f_2-\mu rk_2-w_2l_2+g-\sigma(k_2+w_2l_2)\right]}_{\text{top-up tax}}-\frac{\delta}{2}g^2$$
$$=(1-t_1^N)(f_1-\mu rk_1-w_1l_1-g)-(1-\mu)rk_1+(1-t_m)(f_2-\mu rk_2-w_2l_2+g)$$
$$-(1-\mu)rk_2+\underbrace{\sigma(t_m-t_2^N)(k_2+w_2l_2)}_{\text{tax saved from the SBIE}}-\frac{\delta}{2}g^2.$$

Under the QDMTT, the short-run revenues of two countries are:

$$R_1^m = t_1^N(f_1-\mu rk_1-w_1l_1-g),$$

$$R_2^m = t_2^N(f_2-\mu rk_2-w_2l_2+g)+(t_m-t_2^N)[f_2-\mu rk_2-w_2l_2+g-\sigma(k_2+w_2l_2)]$$
$$= t_m(f_2-\mu rk_2-w_2l_2+g)-\underbrace{\sigma(t_m-t_2^N)(k_2+w_2l_2)}_{\text{loss from the deduction of SBIE}}.$$

The short-run revenue effect of the GMT is given by the following:

**Proposition E.1.** *In the short run where countries' tax rates are fixed,*

*(i) the large country benefits from the GMT;*

*(ii) introducing a GMT with minimum rate marginally higher than the small country's equilibrium tax without the GMT increases (reduces) the small country's tax revenue if* $\phi_2(t_2^N) := \left[-\frac{\partial k_2/k_2}{\partial t_2/t_2}-\frac{t_2}{1-t_2}\frac{-\bar{l}_2f''_{2lk}}{k_2f''_{2kk}}-\frac{1}{1-t_2}\frac{w_2\bar{l}_2}{k_2}-1\right]\Big|_{t_2=t_2^N}>0\ (<0).$

This proposition holds for a general form of production function. $\phi_2(t_2^N)$ incorporates several key variables evaluated at country 2's tax rate $t_2^N$: the tax elasticity of capital investment (in absolute value) $-\frac{\partial k_2/k_2}{\partial t_2/t_2}$, the elasticity of substitution between labor and capital (in absolute value) $-\frac{\bar{l}_2f''_{2lk}}{k_2f''_{2kk}}$, and the ratio of payroll cost to capital stock $\frac{w_2\bar{l}_2}{k_2}$.

Notably, Proposition E.1(ii) also applies to the baseline model where labor input is absent. Letting $\bar{l}_2=0$, we have $\phi_2(t_2^N)=-\frac{\partial k_2^N/k_2^N}{\partial t_2^N/t_2^N}-1$. So when capital is the only (variable) production input, a marginal GMT reform increases (reduces) the short-run revenue of country 2 if the elasticity of capital investment in absolute value is greater (smaller) than unity. This is an equivalent statement of Proposition 1(ii) in the baseline model.

Given the Cobb-Douglas production technology, we can get that: $\forall i = 1, 2$, $\phi_i(t) = \left[-\frac{\partial k_i/k_i}{\partial t_i/t_i} - \frac{t_i}{1-t_i}\frac{-\bar{l}_i f''_{ilk}}{k_i f''_{ikk}} - \frac{1}{1-t_i}\frac{w_i\bar{l}_i}{k_i} - 1\right]\Big|_{t_i=t} = \frac{t(1-\mu-\beta(1-\mu t))}{(1-\lambda)(1-t)(1-\mu t)} - \frac{\beta r(1-\mu t)}{\lambda(1-t)^2} - 1 =: \phi(t)$.

In the long run, both the MNE and the governments can adjust their behavior in response to the GMT. As in the baseline model, denote by $t_i(t_j)$ country $i$'s best response function absent the GMT,[43] and the Nash equilibrium tax rate of country $i$ under the GMT is denoted by $t_i^m$. We can establish the following:

**Proposition E.2.** *After the introduction of the GMT, two countries set equilibrium tax rates in the following way:*

*(i) For $\phi(t_m) \leqslant 0$, $t_1^m = t_1(t_m)$, $t_2^m = t_m$;*

*(ii) For $\phi(t_m) > 0$, two countries' equilibrium taxes are $(t_1^m = t_1(t_m), t_2^m = \breve{t}_2)$ when $R_1(t_1(t_m), t_m) > R_1^m(\breve{t}_1, \breve{t}_2)$, and $(t_1^m = \breve{t}_1, t_2^m = \breve{t}_2)$ when $R_1(t_1(t_m), t_m) < R_1^m(\breve{t}_1, \breve{t}_2)$; both $(t_1(t_m), \breve{t}_2)$ and $(\breve{t}_1, \breve{t}_2)$ are equilibrium tax rates when $R_1(t_1(t_m), t_m) = R_1^m(\breve{t}_1, \breve{t}_2)$, where $\breve{t}_i < t_m$, $i = 1, 2$.*

When $\phi(t_m) \leqslant 0$, the GMT binds the small country and makes the large country choose tax rate along the initial best-response function. By contrast, when $\phi(t_m) > 0$, the small country will undercut the GMT rate and collect top-up taxes at equilibrium. Consider introducing a GMT with minimum rate marginally higher than the small country's equilibrium tax without the GMT. Then using Propositions E.1(ii) and E.2(i), we can conclude that: the marginal reform will raise the small country's revenue in the long run if it harms the small country in the short run, as in the baseline model (see Remark 5).

[43] $\forall t_j \in \left(0, \frac{1-\lambda}{1-\mu\lambda}\right)$, $t_i(t_j) \in \left(0, \frac{1-\lambda}{1-\mu\lambda}\right)$ is the unique solution to the equation $r^{-\frac{\lambda}{1-\lambda}}\left(\lambda\bar{l}_i^{\beta}\right)^{\frac{1}{1-\lambda}}\left(\frac{1-t_i}{1-\mu t_i}\right)^{\frac{1}{1-\lambda}}\left\{\frac{(1-\beta)(1-\mu t_i)}{\lambda(1-t_i)} - \frac{t_i(1-\mu)[1-\mu-\beta(1-\mu t_i)]}{(1-\lambda)(1-t_i)^2(1-\mu t_i)} - \mu\right\} - \frac{2t_i - t_j}{\delta} = 0$.